\documentclass[%
 reprint,
 amsmath,amssymb,
 aps,
prstab,
floatfix,
]{revtex4-1}

\usepackage[table]{xcolor}
\usepackage{graphicx}
\usepackage{pict2e}

\usepackage{dcolumn}
\usepackage{bm}
\usepackage[hidelinks]{hyperref} 
\usepackage[all]{hypcap}

\usepackage{relsize}

\usepackage{booktabs}

\usepackage{mathtools}

\newcommand{\enquote}[1]{``#1''} 

\usepackage{xspace}
\newcommand{\unit}[1]{\ensuremath{\,\mathrm{#1}}}
\newcommand{\cesrta}{{C{\smaller[2]ESR}TA}\xspace}

\newcommand{\ev}{{\rm e}\kern-1.pt{\rm V}}
\newcommand{\gev}{{\rm Ge}\kern-1.pt{\rm V}}
\newcommand{\mev}{{\rm Me}\kern-1.pt{\rm V}}
\newcommand{\kev}{{\rm ke}\kern-1.pt{\rm V}}
\newcommand{\tev}{{\rm Te}\kern-1.pt{\rm V}}
\newcommand{\gevsq}{\mbox{$\mathrm{{\rm Ge}\kern-1.pt{\rm V}}^2$}}

\newcommand{\ud}{{\rm d}}

\def\lsim{\mathrel{\rlap{\lower4pt\hbox{\hskip1pt$\sim$}}
    \raise2pt\hbox{$<$}}} 
\def\gsim{\mathrel{\rlap{\lower4pt\hbox{\hskip1pt$\sim$}}
    \raise2pt\hbox{$>$}}} 

\begin{document}

\title
{
  Measurement and modeling of electron-cloud-induced betatron tune shifts at the
  Cornell Electron-positron Storage Ring test accelerator
}

\author{S.~Poprocki}
\email{E-mail: stp44@cornell.edu}
\author{S.\,W.~Buechele}
\author{J.\,A.~Crittenden}
\author{K.~Rowan}
\author{D.\,L.~Rubin}
\author{J.\,E.~San~Soucie}
\affiliation
{
CLASSE, Cornell University, Ithaca, NY 14853, USA
}

\date{\today} 

\begin{abstract}
We report on extensive measurements at the Cornell Electron-positron Storage Ring of electron-cloud-induced betatron tune shifts
for trains of positron bunches at 2.1 and 5.3\unit{GeV} with bunch populations ranging between \mbox{$0.64 \times 10^{10}$} and
\mbox{$9.6 \times 10^{10}$}.
Measurements using a witness bunch with variable distance from the end of the train and variable bunch population
provide information on
cloud decay and cloud pinching during the bunch passage.
We employ Monte Carlo simulations of the reflection and absorption of synchrotron radiation photons 
to determine the pattern of absorption sites around the circumference of the storage ring.
The Geant4 simulation toolkit is used to model the interactions of the photons with the beampipe wall
and determine the production energy and location distributions of the photoelectrons
which seed the electron cloud.
An electron cloud buildup model based on fitted
ring-averaged secondary-yield properties of the vacuum chamber predicts tune shifts in
good agreement with the measurements.
\end{abstract}

\pacs{Valid PACS appear here}
\maketitle


%

\section{Introduction}

The buildup of low-energy electron densities in the vacuum chamber of a positron storage ring can result in 
betatron tune shifts, instabilities and emittance growth. We describe techniques to measure
electron-cloud-induced tune shifts, and to use the measurements to constrain
predictive numerical models of electron cloud phenomena. Analytic and  numerical treatments of
electron cloud (EC) contributions to coherent tune shifts were originally presented in Ref.~\cite{APAC01:WEP056}
and further developed in Ref.~\cite{PRSTAB6:081002}.

The Cornell Electron-positron Storage Ring (CESR) was re-configured as a test accelerator in 
2008~\cite{JINST10:P07012}. A comprehensive summary of the project, which included electron-cloud buildup
and low-emittance lattice studies, can be found 
in the {\cesrta} Phase~I Report~\cite{CLNS:12:2084}.
The results reported here concern three lattice configurations of the CESR ring: the test accelerator configurations
at 2.1\unit{\gev} and at 5.3\unit{\gev}, and the 6.0\unit{\gev} upgrade to be commissioned in 2019.
Table~\ref{tab:lattices} lists the parameters of these three lattice configurations.
\begin{table}[hbpt]
\centering
\caption{Lattice and beam parameters for the three configurations of the CESR ring addressed in this report:
  the 2.1\unit{\gev} and 5.3\unit{\gev} lattice configurations for which betatron tune shifts were measured and for
  which simulations were performed, and the 6.0\unit{\gev} configuration for which the
  model was used to assess effects of electron cloud buildup on performance.
        }
\label{tab:lattices}
\linethickness{3mm}
\renewcommand{\arraystretch}{1.1} 
\begin{tabular}{lccc}
\hline
\hline
Beam energy (\gev) & 2.085 & 5.289 & 6.000\\
Circumference (m) & \multicolumn{3}{c}{768.44}\\
Bunch current (mA/bunch)& 0.4--0.7 & 2.0--6.0 & 2.2--4.4 \\
Number of bunches & 30 & 20 & 45--90\\
Beam current (mA) & 12--21 & 40--120 & 200\\
RF frequency (MHz) & \multicolumn{3}{c}{500}\\
Energy loss per turn (\mev) & 0.19 & 1.1 & 1.8 \\
Momentum compaction ($10^{-3}$) & 6.7 & 9.2 & 5.7\\
Bunch length (mm) & 9.2 & 15.8 & 15.6\\
Bunch spacing (ns) & 14 & 14 & 14\\
Energy spread ($10^{-4}$) & 8.1 & 6.5 & 7.6 \\
Horizontal tune & 14.5639 & 11.2853 & 16.545\\
Vertical tune & 9.5984 & 8.7914 & 12.63\\
Synchrotron tune & 0.07354 & 0.04623 & 0.03416\\
Horizontal emittance (nm) & 3.2 & 97 & 30 \\
Vertical emittance (nm) & 0.035 & 1 & 0.1\\
\hline
\hline
\end{tabular}
\renewcommand{\arraystretch}{1.0}
\end{table}

In Sec.~\ref{sec:meas_method} below we discuss and compare methods of measuring bunch-by-bunch
betatron tune shifts.
A comprehensive set of measurements along trains of positron bunches at 2.1\unit{\gev} and
5.3\unit{\gev} is shown.
We describe in Sec.~\ref{sec:sim_method} the full procedure of electron cloud simulation
starting with the 
generation of photons from synchrotron radiation, tracking of the photons in a 3D model of
the vacuum chamber including 
reflections and absorption of the photons, the production of photoelectrons,
the buildup of electron densities along a train of 
bunches, and the calculation of betatron tune shifts.

Although electron cloud buildup models have been successful in simulating tune shifts~\cite{IPAC10:TUPD024,PAC11:WEP108,ECLOUD12:Wed1000,NAPAC16:TUPOB23}
and vertical emittance growth~\cite{IPAC16:TUPOR021,NAPAC16:WEA2CO03} 
in general agreement with measurements, their predictive power has been limited.
Furthermore, no
model
has yet reproduced measurements of
horizontal and vertical tune shifts over as wide a range of bunch population and beam energy as 
considered in this analysis.

Models of electron cloud formation, which are the basis for prediction of tune shifts and emittance growth,
typically depend on phenomenological descriptions of much of the underlying physics. The model parameters are tuned
so that simulations based on the model are consistent with measurements. 
In an effort to improve the predictive power of the model, 
we replace the phenomenological descriptions with first-principles calculations for
two of the processes critical to the determination
of cloud growth.
We employ the Synrad3D~\cite{PRAB20:020708} code to calculate azimuthal location distributions of photons absorbed
on the vacuum chamber walls, including
their energies and angles of incidence throughout the circumference of the
CESR ring. These calculations include photon reflectivity and the effect of surface roughness.
The Geant4~\cite{AGOSTINELLI2003250} simulation tookit is then used to calculate the emission of photoelectrons by the
absorbed photons
into the vacuum chamber volume~\cite{CLASSE:REU2017:buechele}. 
The description of the physics of secondary-electron yield (SEY) remains phenomenological. Parameters of the secondary-yield model
are fit to the large dataset of betatron tune shift measurements
collected at CESR.

Section~\ref{sec:sim_result} discusses these results and draws conclusions about how
betatron tunes respond to various cloud buildup characteristics. Finally, the model
with the best-fit SEY parameters
is applied to obtain estimates for the consequences of electron cloud buildup for operation
of the major CESR upgrade (CHESS-U) to be completed in 2019~\cite{PhysRevAccelBeams.22.021602}.

%
\section{Tune Shift Measurement}
\label{sec:meas_method}
Tune shifts have been measured in a number of ways at {\cesrta}.
A relatively straightforward technique is to kick the entire train all at once with a single-turn pinger, and then record
turn-by-turn position data for each bunch.
In the limit where the positron bunch is oscillating transversely on passage through a static electron cloud,
an FFT of the position data yields the betatron tune~\cite{CLNS:12:2084,IPAC10:TUPD024} and the shift due to the presence of the cloud.
%
%
But because the cloud follows the pinger-induced horizontal motion of the train,
the measurement of horizontal tune shifts by this method is difficult to interpret.
In general,
low-energy electrons emitted from the top (bottom) of the vacuum chamber
are accelerated by the positron bunch and strike the bottom (top) of the chamber.  In the dipole magnets,
the resulting secondaries are trapped by the magnetic field lines  in
a vertical band of width comparable to that of the bunch. 
A horizontal ping, with pulse length long compared to the train length,
moves the bunch train coherently, and thus the cloud as well.
This measurement technique is thus insensitive to horizontal tune shifts, since the
test bunch receives no coherent kick from the co-moving cloud.
A further limitation of the technique is that the presence of multiple peaks in the FFT from
coupled-bunch motion contaminates the signal.

Better results are obtained by enabling the bunch-by-bunch feedback, and disabling it
one bunch at a time to measure the tune of that bunch.
The self-excitation (no external kick applied) typically yields a measurable signal, but the precision of the measurement
is improved by kicking the single bunch with a gated strip-line kicker.
This technique is further refined by driving the bunch with a tune
tracker~\cite{PAC11:MOP215}.
The tune tracker phase-locks an  oscillator to the
observed betatron signal, providing a frequency source for coherent excitation
of steady-state betatron motion.  It also provides a digital clock that is
synchronized to the instantaneous betatron motion, which can be used for synchronous
detection of betatron signals.
The measurements
with the tune tracker were done separately in horizontal and vertical planes.
Betatron tunes were measured along the train, one bunch at a time, and bunch currents were monitored and topped off between measurements as needed---typically after every 1--5 bunches.

\begin{figure}[b]
\centering
\includegraphics[width=0.99\columnwidth]{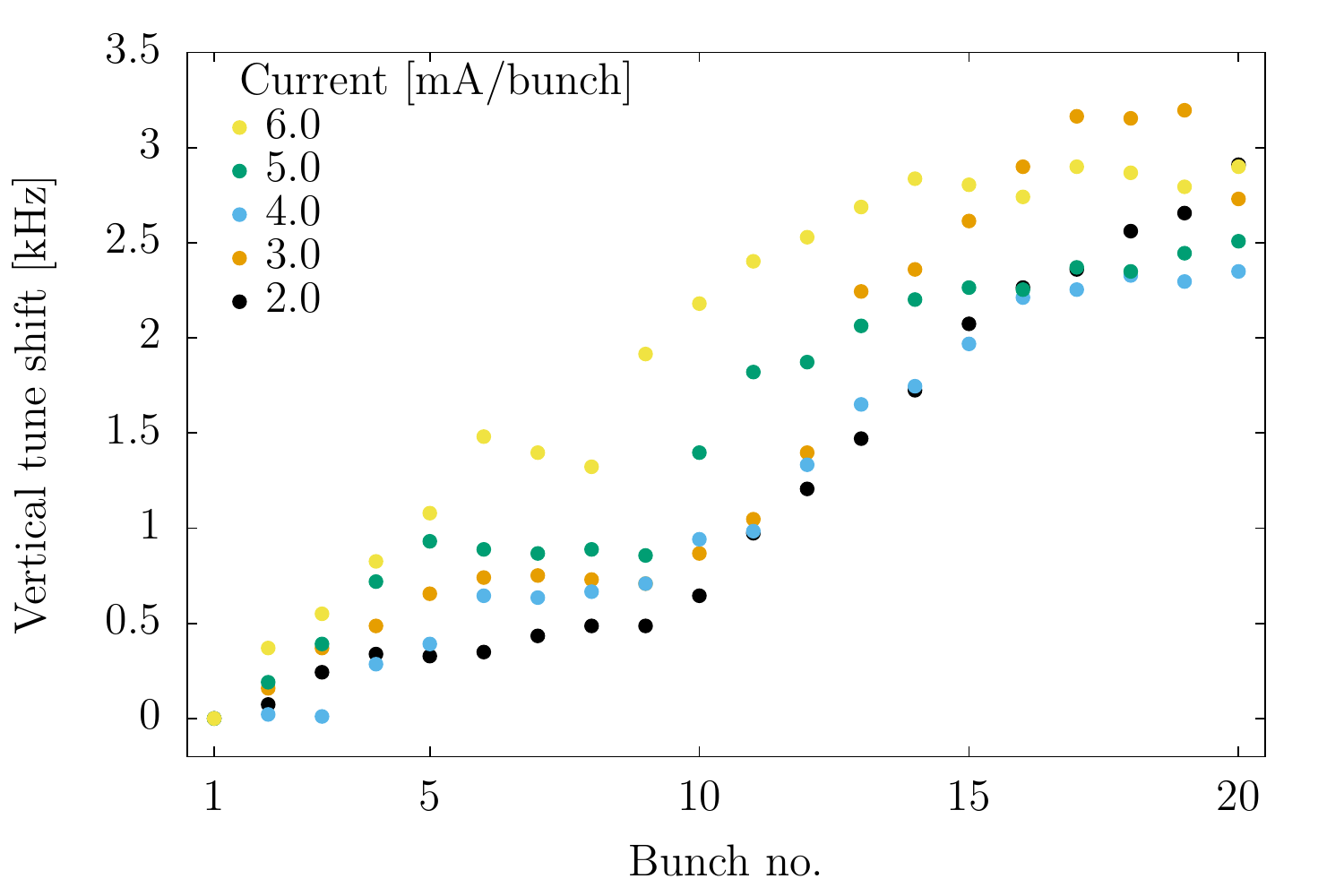}
   \caption{
   Vertical betatron tune shifts measured using the ``pinging'' method
   along a 20-bunch train of positrons at 5.3\unit{\gev} for values of the bunch current ranging
   from~2~to~6\unit{mA/bunch}
   (3.2--$9.6\times10^{10}$ bunch populations).
   }
   \label{fig:tunes_ping}
\end{figure}

\begin{figure}[tb]
\centering
\includegraphics[height=4.2in]{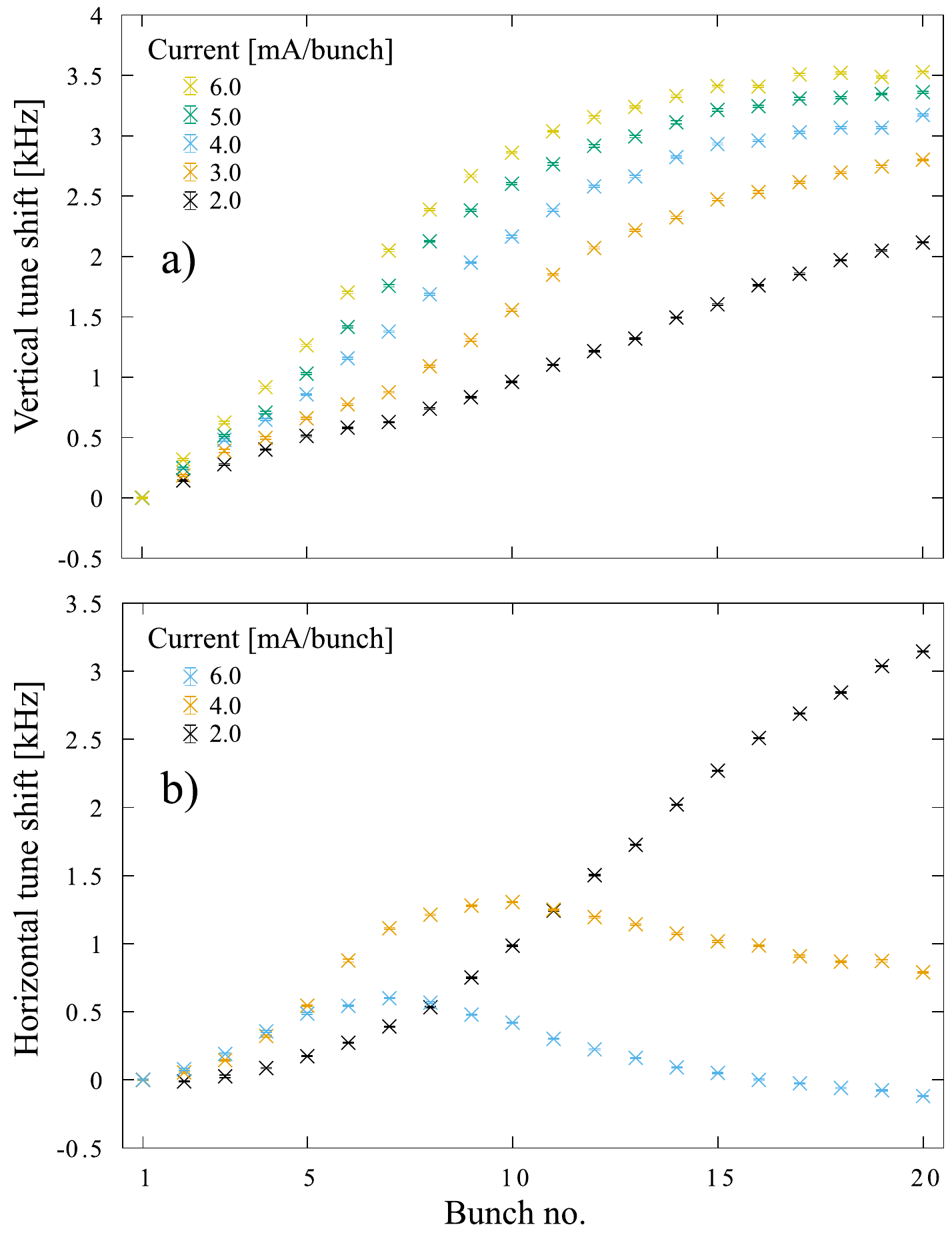}
\caption{
  Tune shifts measured in the a)~vertical and b)~horizontal planes
    using the tune tracker for a 20-bunch train of positrons with values for the bunch current
  ranging between~2~and~6\unit{mA/bunch} 
   (3.2--$9.6\times10^{10}$ bunch populations) at 5.3\unit{\gev}. 
   Data were recorded separately for each of the two planes.
      }
   \label{fig:5gev_tuneshift_meas}
\end{figure}

Tune shifts measured using the pinging method for 20-bunch trains of positrons at 5.3\unit{\gev}
and for several values of the bunch current are shown in Fig.~\ref{fig:tunes_ping}.
The bunch spacing is 14\unit{ns}. For the CESR
revolution frequency of 390\unit{kHz}, a tune shift of 1\unit{kHz} corresponds to a fractional tune shift $\Delta Q = 0.0026$. 
Large bunch-to-bunch fluctuations as well as overlap of data are observed.
The tune shift measurements obtained using the tune tracker are shown in Fig.~\ref{fig:5gev_tuneshift_meas},
and exhibit vertical tune shifts increasing monotonically with bunch current.
These measurements are more useful than those obtained via the pinging method
since a)~the single bunch tune measurement using the tune tracker is more accurate, and b)~all of the other bunches
in the train are stabilized via feedback, thus eliminating coupled bunch motion.

The horizontal tune shift reaches a maximum along the train, decreasing for later bunches. The maximum
tune shift occurs earlier in the train as bunch current increases.
This behavior is understood in terms of the ``cloud splitting'' effect in dipole magnetic fields. 
The cloud electron energies resulting from the attractive kick imparted by passage of the positron bunch
increase with bunch current. The electron trajectories are pinned along the vertical field lines in a tight spiraling motion.
The secondary-emission yield has a strong dependence on the incident electron energy, with peak yield at an energy of a few hundred eV.
(For a description of secondary-electron emission processes, see, for example, Ref.~\cite{HBruining1954:PhysApplSEE}.)
The electrons near the positron bunch in the vertical plane containing the beam are accelerated to the highest energies, so
these are the first to strike the vacuum chamber wall at energies exceeding the maximum in the SEY energy dependence.
As the cloud builds up during the passage of the train,
the location on the beampipe wall of maximum average SEY moves away from the vertical plane
containing beam and
the dense vertical stripe of the cloud first widens and then splits into two stripes.
%

\begin{figure}[thb]
\centering
\includegraphics[height=4.2in]{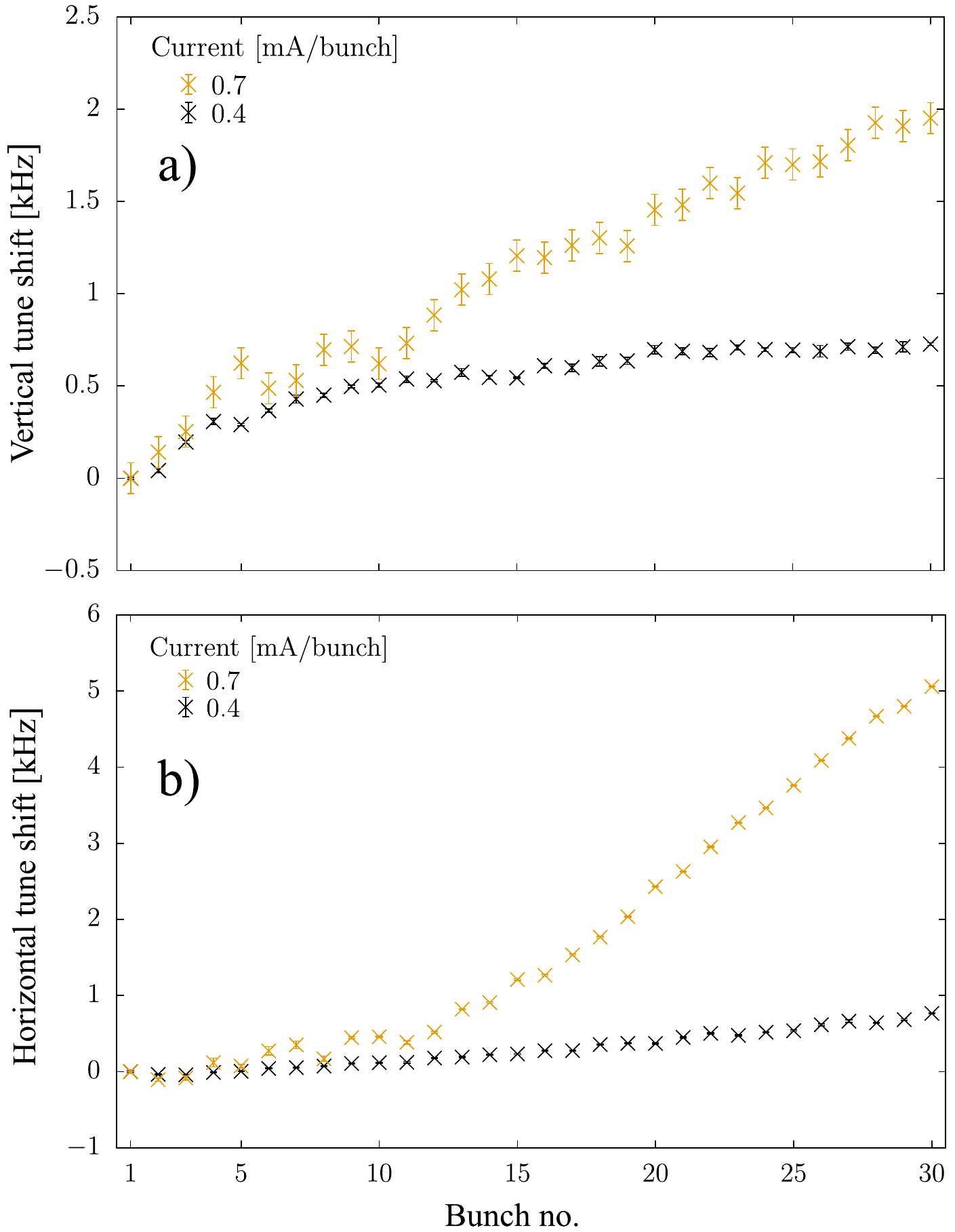}
   \caption{
   Tune shifts measured in the a)~vertical and b)~horizontal planes using the tune tracker
   for a 30-bunch train of positrons at 0.4 and 0.7\unit{mA/bunch} 
   ($0.64\times10^{10}$ and $1.12\times10^{10}$ bunch populations) at 2.1\unit{\gev}.
   }
   \label{fig:2gev_tuneshift_meas}
\end{figure}

Tune shift measurements taken with the tune tracker for positrons at 2.1\unit{\gev} are
shown in Fig.~\ref{fig:2gev_tuneshift_meas}.
The fluctuations and larger uncertainties observed in the vertical tune shift
measurements at 0.7\unit{mA/bunch} were reduced in subsequent measurements at
the other bunch currents by averaging over measurements collected at an increased acquisition rate.
The horizontal tune shift depends on the bunch current in a non-linear way,
increasing by more than a factor of five as the the bunch current increases
from \mbox{0.4 to 0.7\unit{mA/bunch}}.
The roughly linear increase in tune shift with bunch number,
beginning with bunch 11, is
shown by the modeling to be characteristic of the cloud growth in the
dipoles (see Sec.~\ref{sec:sim_result}).
Note the very different bunch currents in the measurements at 2.1\unit{\gev} and 5.3\unit{\gev}
and the nonlinear dependence of tuneshift on bunch current and beam energy.
While the synchrotron photon emission
rate increases linearly with beam energy and bunch current, the higher beam kicks result in
cloud electron energy distributions which span the maximum in the dependence of SEY on incident
electron energy, leading to saturation. 

%

\section{Simulation Method\label{sec:sim_method}}
The modeling of electron cloud effects on beam dynamics proceeds in four steps: 
1)~3D calculation of the pattern of absorbed synchrotron radiation around 
the ring including the effects of photon reflections~\cite{PRAB20:020708}, 
2)~simulation of the interactions of absorbed photons with the vacuum chamber wall
which lead to the emission of
electrons~\cite{ALLISON2016186,1742-6596-664-7-072021,AGOSTINELLI2003250},
3)~a~time-sliced weak-strong model~\cite{CERN:SL2002:016AP,ECLOUD12:Fri1240} for electron
cloud development
along a train of positron bunches, 
including a phenomenological model for SEY from the beampipe walls, and
%
4)~calculations of betatron tune shifts using the space-charge electric field gradients
derived from the cloud buildup model~\cite{CLNS:12:2084,IPAC10:TUPD024,ECLOUD10:PST10,PAC11:WEP108}.
The physics of SEY was parameterized
as described in~Ref.~\cite{PRSTAB5:124404} and the parameters were fit to the tune shift measurements
using an iterative optimization procedure. 
Note that the SEY parameters are the only free parameters in the simulation.
These four steps are described in
Sects.~\ref{sec:sim_method:synrad3d}--\ref{sec:sim_method:sey_optimization} below.

\subsection{Synchrotron radiation photon tracking\label{sec:sim_method:synrad3d}}
\begin{figure}[tbhp]
\centering
\includegraphics[width=\columnwidth]{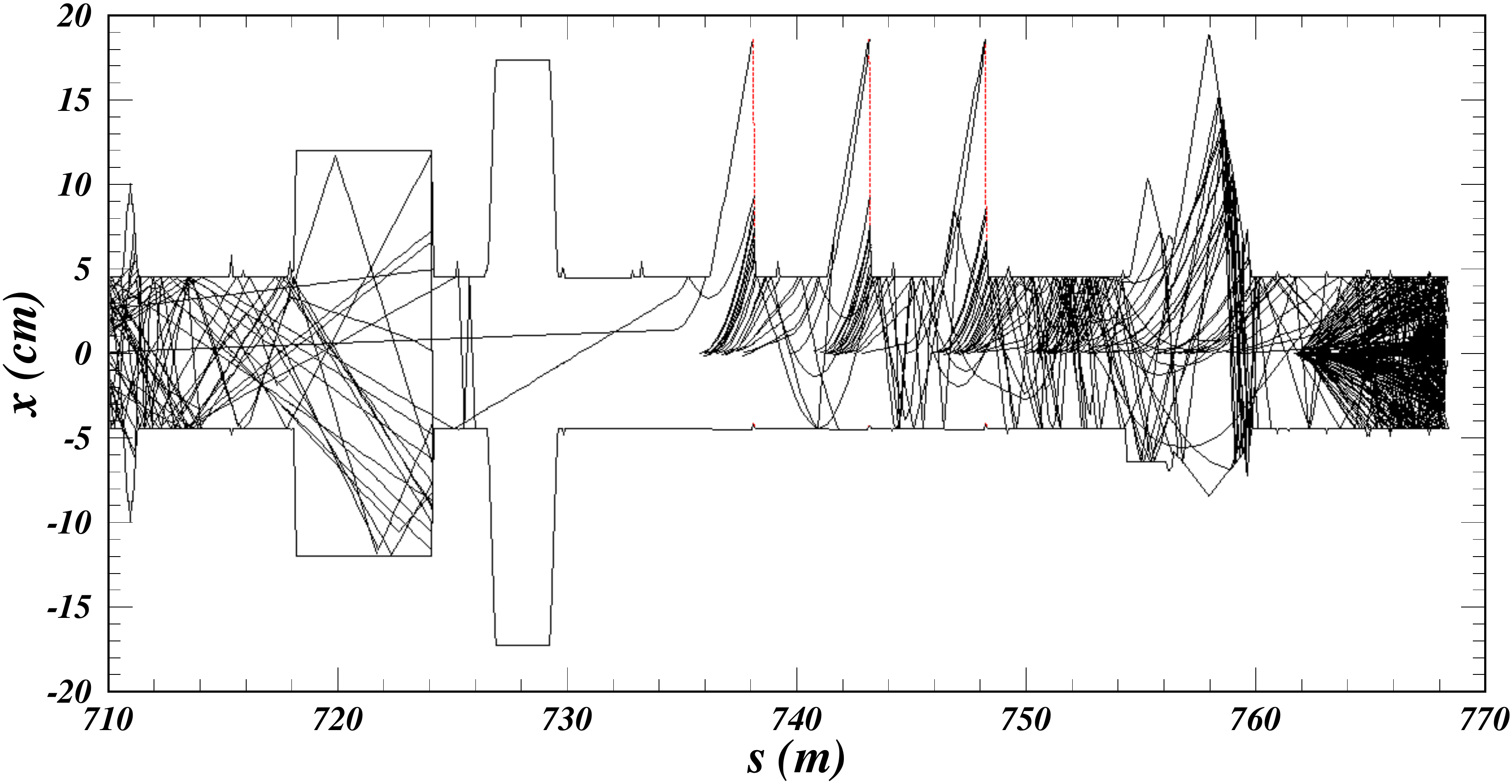}
\caption{
  Modeled photon trajectories in a section of the CESR ring which includes various vacuum system
  components as well as X-ray beam exit ports. Projections of the horizontal position $x$
  of the photons
  relative to the positron reference orbit are shown.
  The positrons travel in the direction of increasing $s$
  coordinate. The red vertical lines represent the
  exit port windows; any photon hitting those surfaces are terminated and excluded from
  the absorbed-photon rate.
}
   \label{fig:photon_trajectories}
\end{figure}

An essential tool in this study is the photon-tracking code Synrad3D~\cite{PRAB20:020708}, which simulates the generation of individual photons
radiated by the positron beam, and incorporates a user-defined 3D model of the vacuum chamber to model the reflection and absorption of photons
using the Bmad library~\cite{NIMA558:356to359} and X-ray data from an LBNL database~\cite{ADNDT54:181to342}. Figure~\ref{fig:photon_trajectories}
shows a plan view of photon trajectories in a region of the CESR ring which includes X-ray beamline exit windows at which incident photons are
not included in the tally of photons absorbed in the vacuum chamber walls.

Photon reflectivity plays a crucial role in electron cloud buildup, since it determines the distribution of photon
absorption sites around the ring. Furthermore, without photon reflectivity, few photons could be absorbed on the top and bottom
of the beampipe, where photoelectron production is the primary source of cloud generation in the vertical plane
containing the beam, which is particularly important in dipole magnetic fields.

\begin{figure}[t]
\centering
\includegraphics[width=\columnwidth]{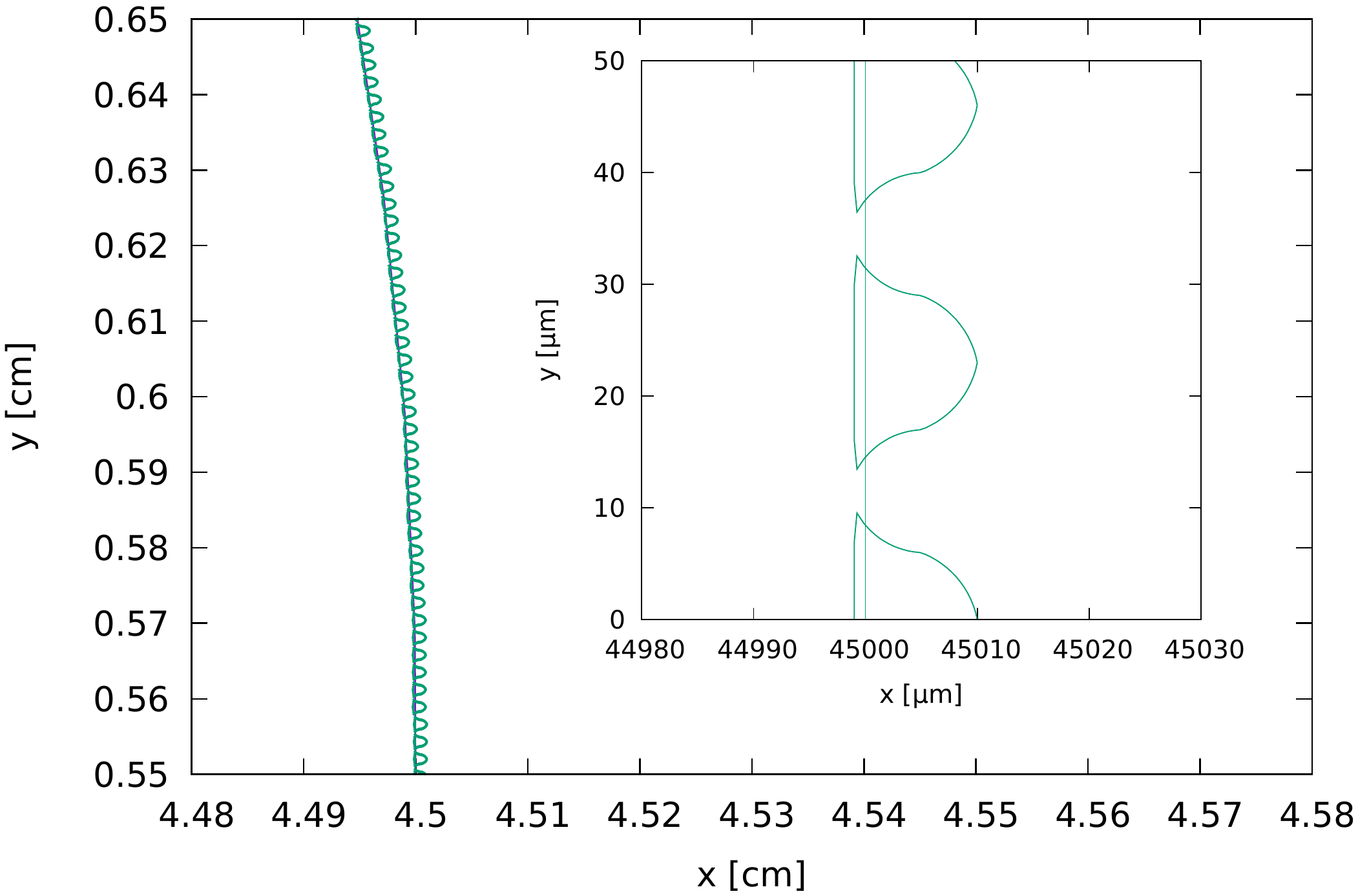}
   \caption{Schematic diagram of the 10-$\mu$m-deep grooves on the CESR vacuum chamber wall used in the photon reflectivity model.
     The simulated vacuum chamber is the union of geometric shapes.
           }
   \label{fig:grooves}
\end{figure}

\begin{figure}[tbp]
\centering
\includegraphics[width=.85\columnwidth]{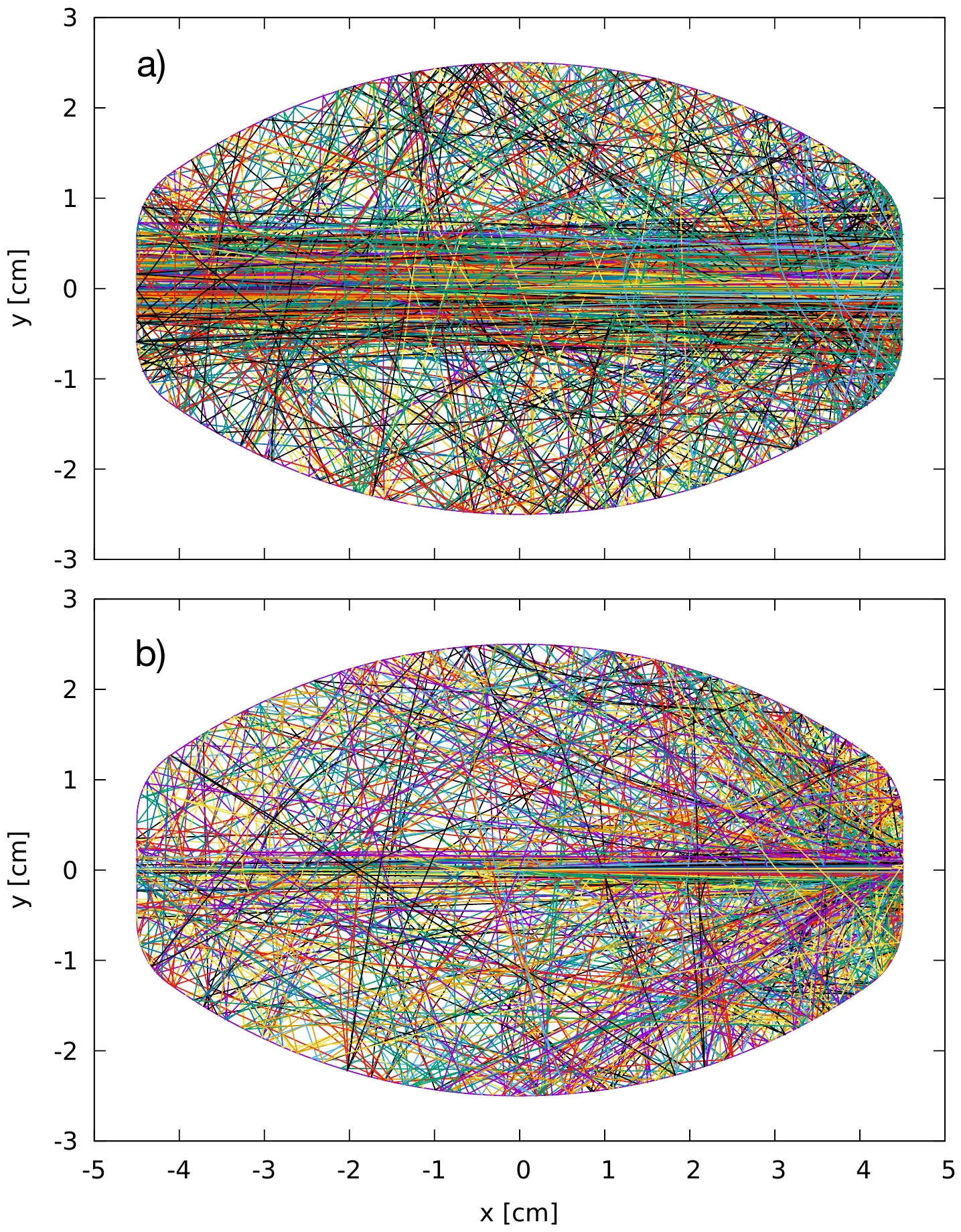}
\caption{Examples of photon trajectories a)~without the groove pattern on the vacuum chamber wall,
  and b)~with grooves.
  The groove pattern results in significantly enhanced scattering out of the horizontal mid-plane.
  The apparent curvature in the tracks is a consequence of the longitudinal bend in the reference
  trajectory in the dipole. (Beam energy is 5.3\unit{\gev}.)
  }
   \label{fig:grooves_photon_tracking}
\end{figure}

A micro-groove structure on the surface of the CESR vacuum chamber
has been measured using atomic force microscopy and
studied in X-ray beams~\cite{PRSTAB18:040704}.
These grooves are roughly parallel to the beam axis and are understood to be caused by the
beampipe extrusion process.
Their effect is accounted for
%
by incorporating the groove structure into the beampipe model 
and simulating specular reflections in the grooves.
Figure~\ref{fig:grooves} shows a diagram of the modeled grooves used in the photon-tracking
simulation, and Fig.~\ref{fig:grooves_photon_tracking} shows the effect of the grooves on the
photon tracks.

\begin{figure}[tbp]
\centering
\includegraphics[width=\columnwidth]{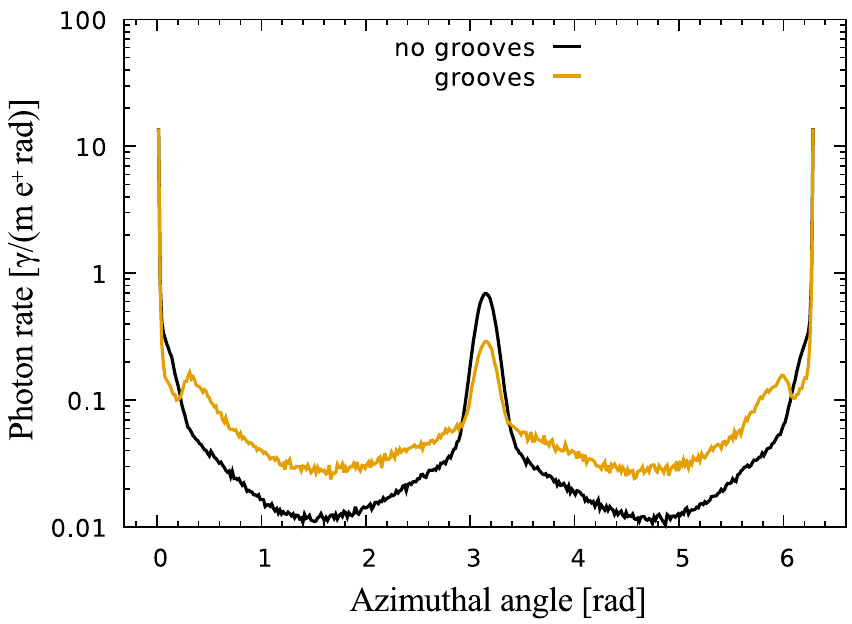}
\caption{Comparison of the azimuthal absorption location of the absorbed photons in the dipole regions when micro-grooves are introduced in the CESR vacuum chamber geometry.
  The azimuthal angle is defined to be $180^\circ$ in the horizontal plane containing the beam axis on the inside of the ring. (Beam energy is 5.3\unit{\gev}.)
   }
   \label{fig:photon_angdist_grooves}
\end{figure}

The transverse absorption location
distribution in Fig.~\ref{fig:photon_angdist_grooves} shows the consequence of the larger reflection angles from grooves in the dipole regions for the case of the 5.3\unit{\gev} beam.
The absorbed photon rate on the top and bottom of the beampipe increases by a factor of about three when the grooves are included.

\begin{figure}[b]
\centering
\includegraphics[width=0.97\columnwidth]{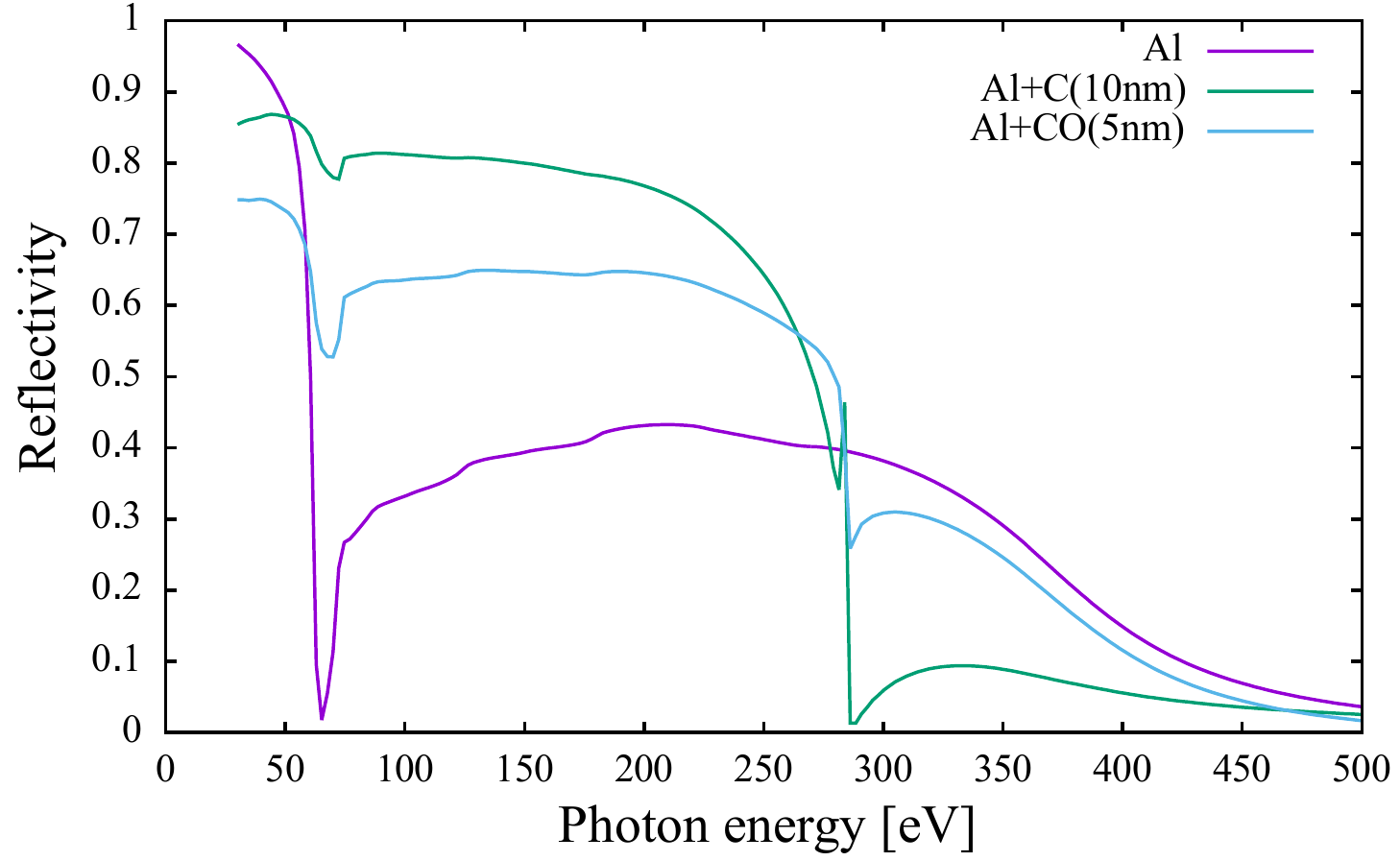}
   \caption{Smooth-surface photon reflectivity versus photon energy for 
   aluminum, aluminum with a \mbox{10-nm} carbon layer, and aluminum with a \mbox{5-nm} 
   carbon monoxide layer, for photons incident at a $5^\circ$ grazing angle. 
   }
   \label{fig:reflectivity}
\end{figure}

The reflectivity is also critically dependent on the material composition of the vacuum chamber wall. Figure~\ref{fig:reflectivity} shows the fraction of photons reflected as a function of photon energy for a $5^\circ$ grazing angle for aluminum with and without C and CO surface layers.
The data were obtained from the LBNL database~\cite{ADNDT54:181to342}.
In validating our modeling studies, we have chosen to use the \mbox{5-nm} CO layer, as motivated in Ref.~\cite{PRSTAB18:040704}.

\begin{figure}[tb]
\centering
\includegraphics[width=0.65\columnwidth]{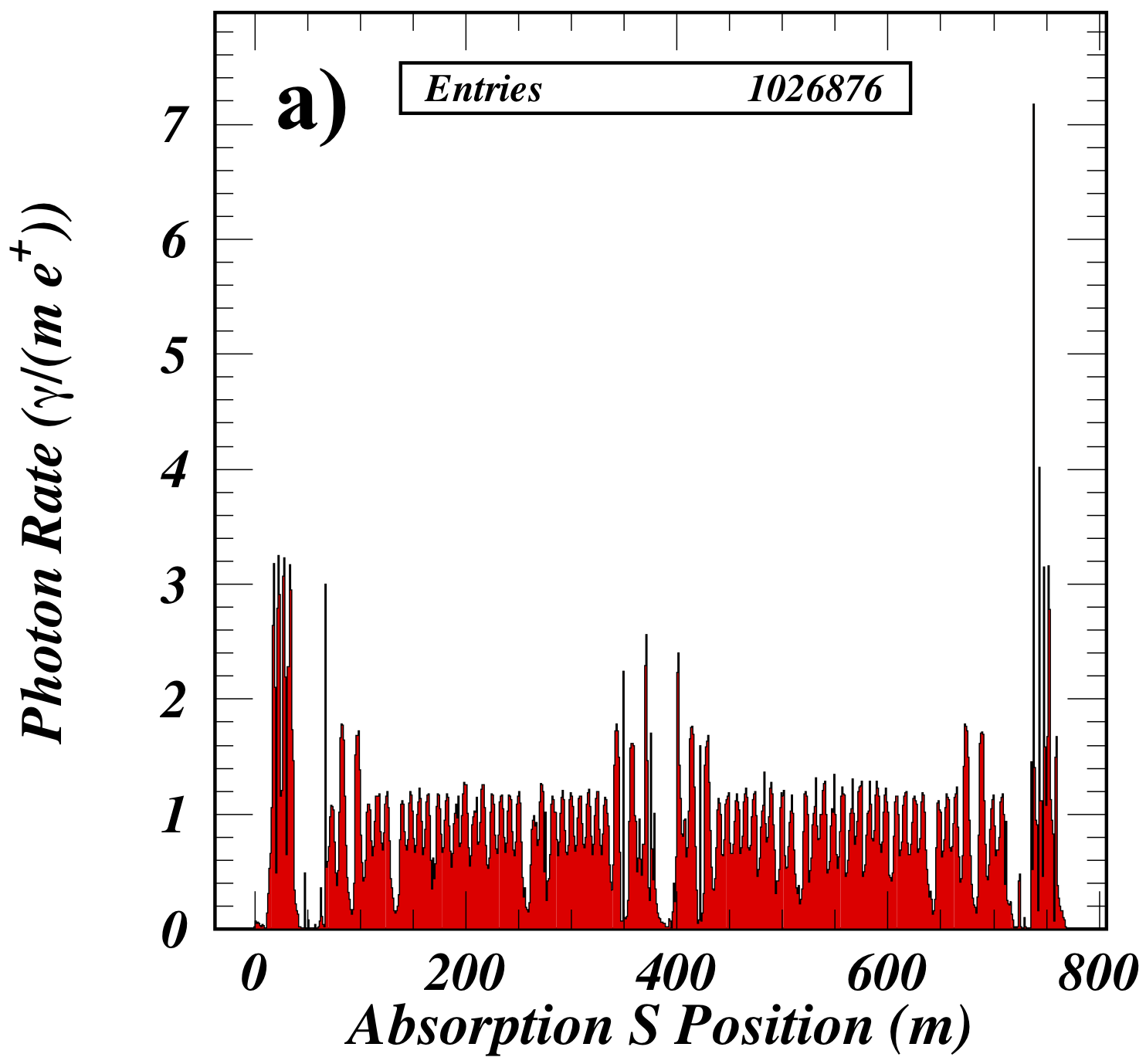}
\includegraphics[width=0.65\columnwidth]{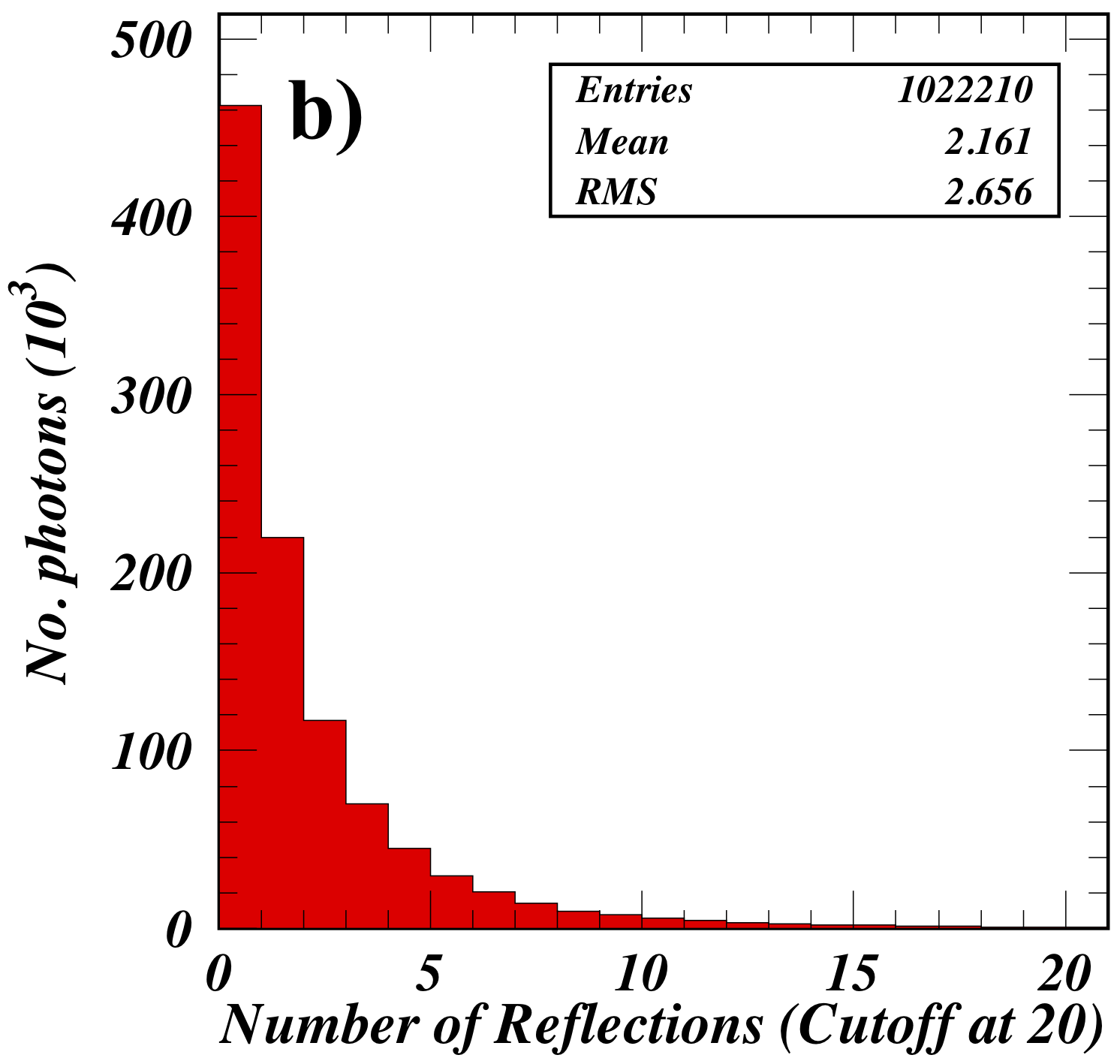}
   \caption{Distributions of absorbed photons in a)~absorption location along the CESR ring, 
      and b)~number of prior reflections. (Beam energy is 5.3\unit{\gev}.)
}
   \label{fig:synrad3d_388_13}
\end{figure}

The photon tracking simulation identifies $10^6$ locations around the CESR ring where photons are absorbed, along with the energy and incident angle of the photon. 
All of the simulation results shown below assume the micro-groove structure, a surface roughness parameter of \mbox{100-nm} RMS for the diffuse component of the scattering,
and a \mbox{5-nm} CO surface layer. The surface roughness parameter value was derived from
the measurements described in Ref.~\cite{PRSTAB18:040704}.
Figure~\ref{fig:synrad3d_388_13}~a)
shows the linear density per beam particle of absorption sites around the 768-m-circumference CESR ring.
The higher densities near the former collider interaction regions at $s=0$, $s=384\unit{m}$, and $s=768\unit{m}$
result from the higher-strength dipole magnets outboard of the straight sections where detectors were formerly installed.
The distribution in the number of reflections prior to absorption is
shown in Fig.~\ref{fig:synrad3d_388_13}~b).  About half of the absorbed photons
are absorbed on the first wall strike. The photon energy is conserved in the modeled reflection process.
However, since the reflection probability is a strong function of
the incident photon energy, the photons absorbed after undergoing a prior reflection are generally 
of lower energy than those absorbed without prior reflections.

\begin{figure}[htbp]
\centering
\includegraphics[width=\columnwidth]{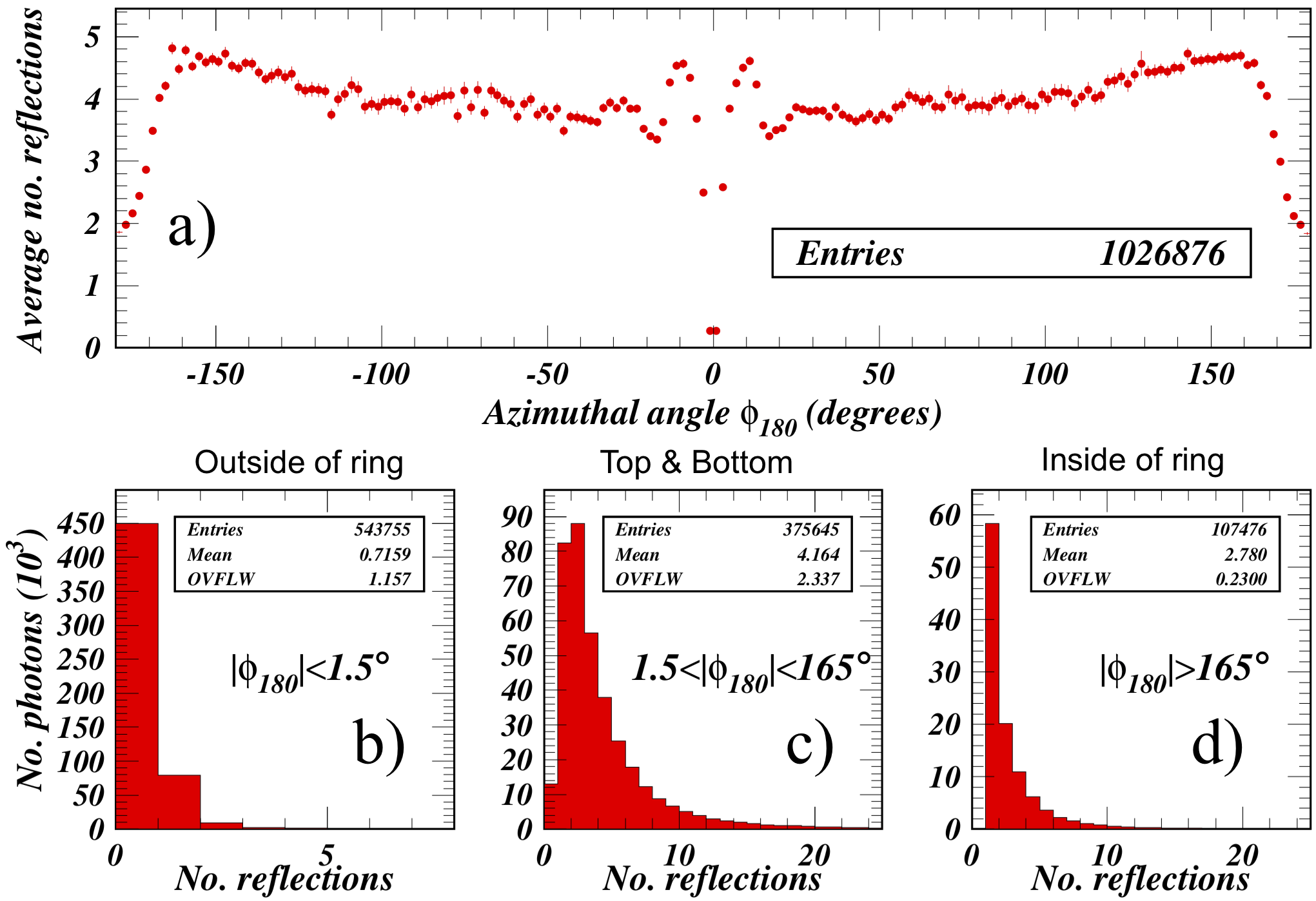}
\caption{Average number of prior reflections for absorbed photons summed over a)~the full ring
  as a function of the azimuthal absorption location on the vacuum chamber wall, $\Phi_{180}$.
  The distributions in the number of reflections are shown for
  the three azimuthal regions \mbox{b)~$|\Phi_{180}|<1.5^\circ$},
  \mbox{c)~$1.5^\circ<|\Phi_{180}|<165^\circ$}, and \mbox{d)~$|\Phi_{180}|>165^\circ$}.
  (Beam energy is 5.3\unit{\gev}.)
}
   \label{fig:photon_refl}
\end{figure}

Only reflected photons strike the top, bottom and inner walls of the vacuum chamber. The typical number of reflections 
before absorption depends on the azimuthal angle $\Phi_{180}$ of the absorption site location, where $\Phi_{180}$ ranges from $-180^\circ$ to $+180^\circ$
with its origin in the mid-plane on the 
outside of the ring.
Figure~\ref{fig:photon_refl}~a)
shows the dependence on this angle of the average number of reflections prior to absorption. 
Note three distinct azimuthal regions. The number of reflections prior to absorption is relatively low on the outer wall (\mbox{$|\Phi_{180}|<1.5^\circ$}), 
since this narrow azimuthal region
  has direct line of sight with the (unreflected) synchrotron radiation. The average number of reflections prior to absorption is roughly constant across the top and bottom of the chamber 
(\mbox{$1.5^\circ<|\Phi_{180}|<165^\circ$}), and it falls again on the inner wall (\mbox{$|\Phi_{180}|>165^\circ$}).
Figures~\ref{fig:photon_refl}~b), c) and d) show the distributions in the number of prior reflections for
the azimuthal ranges \mbox{$|\Phi_{180}|<1.5^\circ$}, \mbox{$1.5^\circ<|\Phi_{180}|<165^\circ$}, and \mbox{$|\Phi_{180}|>165^\circ$}, respectively.
In the region \mbox{$|\Phi_{180}|<1.5^\circ$}, most of the photons (83\%) were not reflected prior to absorption.

\begin{figure*}[t]
\centering
\includegraphics[width=1.5\columnwidth]{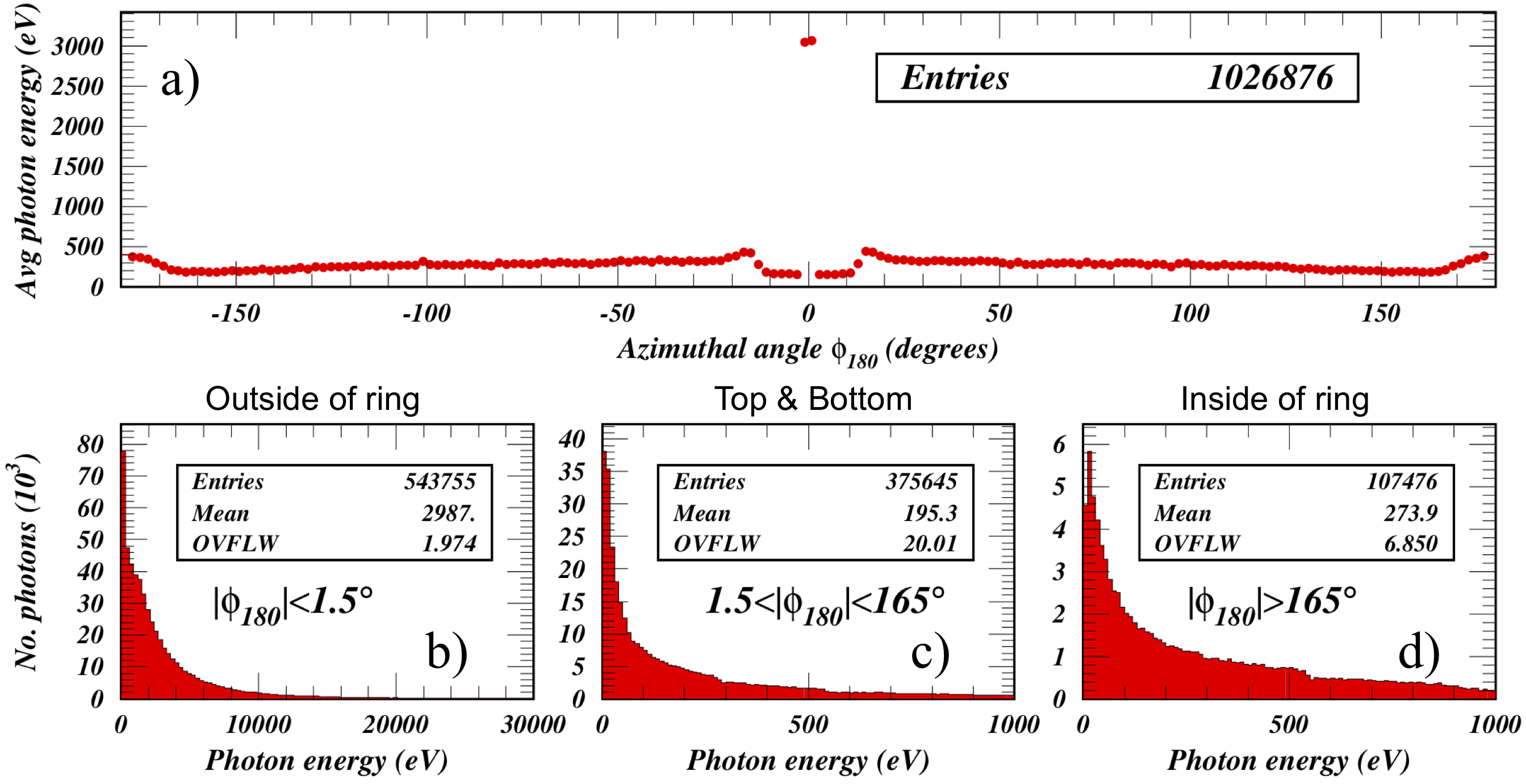}
   \caption{Average energy of the absorbed photons summed over a)~the full ring as a function of the azimuthal absorption location on the vacuum chamber wall,
     $\Phi_{180}$.
     The photon energy distributions
     are also shown for the three azimuthal regions for which electron energy distributions were provided to the electron cloud
     buildup simulation: \mbox{b)~$|\Phi_{180}|<1.5^\circ$},  \mbox{c)~$1.5^\circ<|\Phi_{180}|<165^\circ$},
     and \mbox{d)~$|\Phi_{180}|>165^\circ$}.
     (Beam energy is 5.3\unit{\gev}.)
}
   \label{fig:photon_energies}
\end{figure*}
Due to the correlation of azimuthal angle with number of reflections, and the dependence of the
reflectivity on photon 
energy, one expects a correlation of photon energy with azimuthal angle. 
The dependence of absorbed photon energy on azimuth is shown in detail in
Fig.~\ref{fig:photon_energies}.
Since the probability for electron emission and the energy of the emitted electron depend on photon energy,
and the energy of the absorbed photon depends on azimuthal angle, 
we find that the effective quantum efficiency (that is, the efficiency with which an incident
photon emits an electron) depends strongly on azimuthal angle.

Figures~\ref{fig:photon_energies}~b), c),
and d) illustrate the reason for choosing three distinct azimuthal regions when providing electron energy distributions to the electron cloud buildup simulation, and 
show the average energy of the absorbed photons in the three azimuthal ranges
\mbox{$|\Phi_{180}|<1.5^\circ$},
\mbox{$1.5^\circ<|\Phi_{180}|<165^\circ$} and
\mbox{$|\Phi_{180}|>165^\circ$} is 2987\unit{\ev}, 195\unit{\ev} and 343\unit{\ev}, respectively, averaged over the full ring.

\begin{figure*}[htbp]
\centering
\includegraphics[width=1.5\columnwidth]{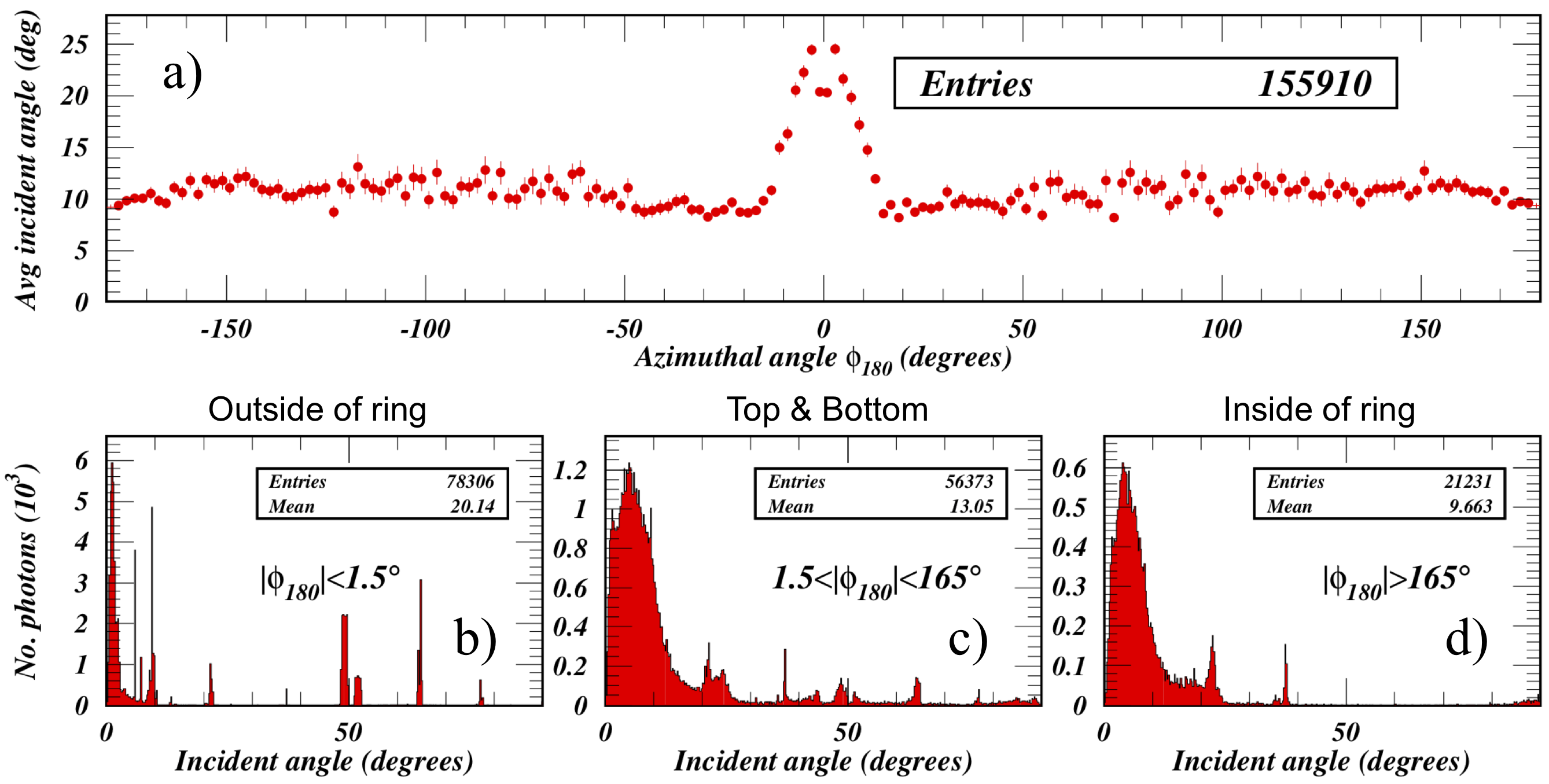}
   \caption{Average angle of incidence (grazing angle) \mbox{$ \langle \theta^{\rm inc}_\gamma \rangle$} of the absorbed photons summed over the field-free regions of the CESR ring a)~as a function of the azimuthal location on the vacuum chamber wall, $\Phi_{180}$. The distributions in three azimuthal regions are shown in \mbox{b)~$|\Phi_{180}|<1.5^\circ$},  \mbox{c)~$1.5^\circ<|\Phi_{180}|<165^\circ$}, and \mbox{d)~$|\Phi_{180}|>165^\circ$}.
   (Beam energy is 5.3\unit{\gev}.)
}
   \label{fig:photon_angles_drifts}
\end{figure*}

\begin{figure*}[htbp]
\centering
\includegraphics[width=1.5\columnwidth]{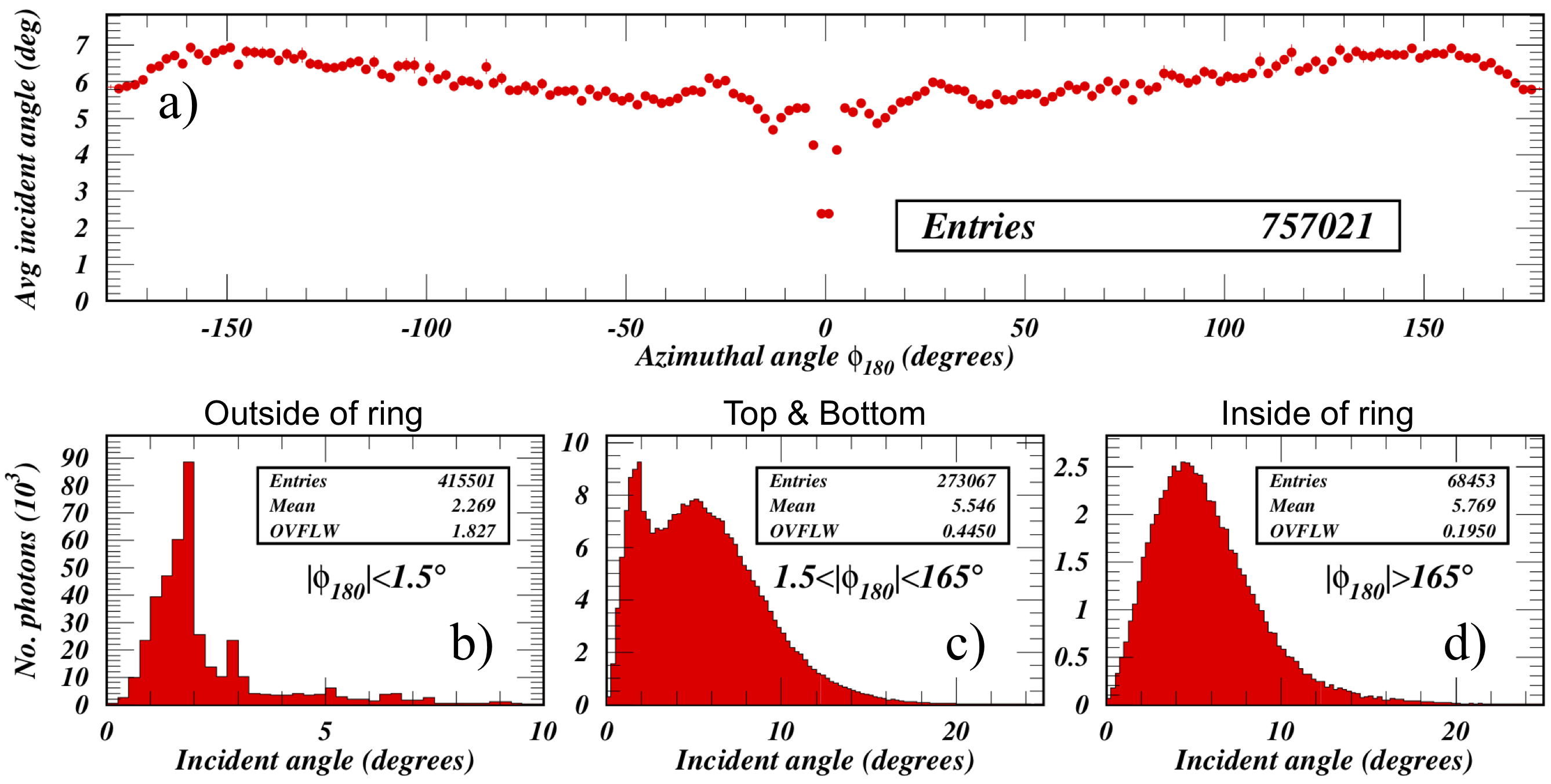}
   \caption{Average angle of incidence (grazing angle) \mbox{$\langle \theta^{\rm inc}_\gamma \rangle$} of the absorbed photons summed over the dipole regions of the CESR ring. These distributions can be compared to those
     summed over the field-free regions of the ring shown in Fig.~\ref{fig:photon_angles_drifts}.
     The dependence of quantum efficiency on incident photon angle results in
     significantly different photoelectron production rates in the field-free and dipole regions. (Beam energy is 5.3\unit{\gev}).
}
   \label{fig:photon_angles_dipoles}
\end{figure*}

We will see below in the section on the Geant4 simulations that the photoelectron production
energy distribution is strongly correlated with the angle of incidence of the photon on the chamber wall. 
Figures~\ref{fig:photon_angles_drifts} and~\ref{fig:photon_angles_dipoles} show details of the photon grazing angle distributions 
as functions of azimuthal impact location, summed over the field-free and dipole regions of the ring, respectively.
The distributions in photon angle of incidence on the vacuum chamber wall are
somewhat
different for the field-free and dipole regions,
with
important
consequences for the average quantum efficiencies.
Generally the photons absorbed in the field-free regions have been multiply reflected and are of lower energy, 
which enhances the quantum efficiency.
The details of the vacuum chamber geometry, such as in gate valves, sliding joints and exit windows, result in a complicated pattern of
photon incident angles around the ring. 

\begin{figure}[b]
\centering
\includegraphics[width=0.8\columnwidth]{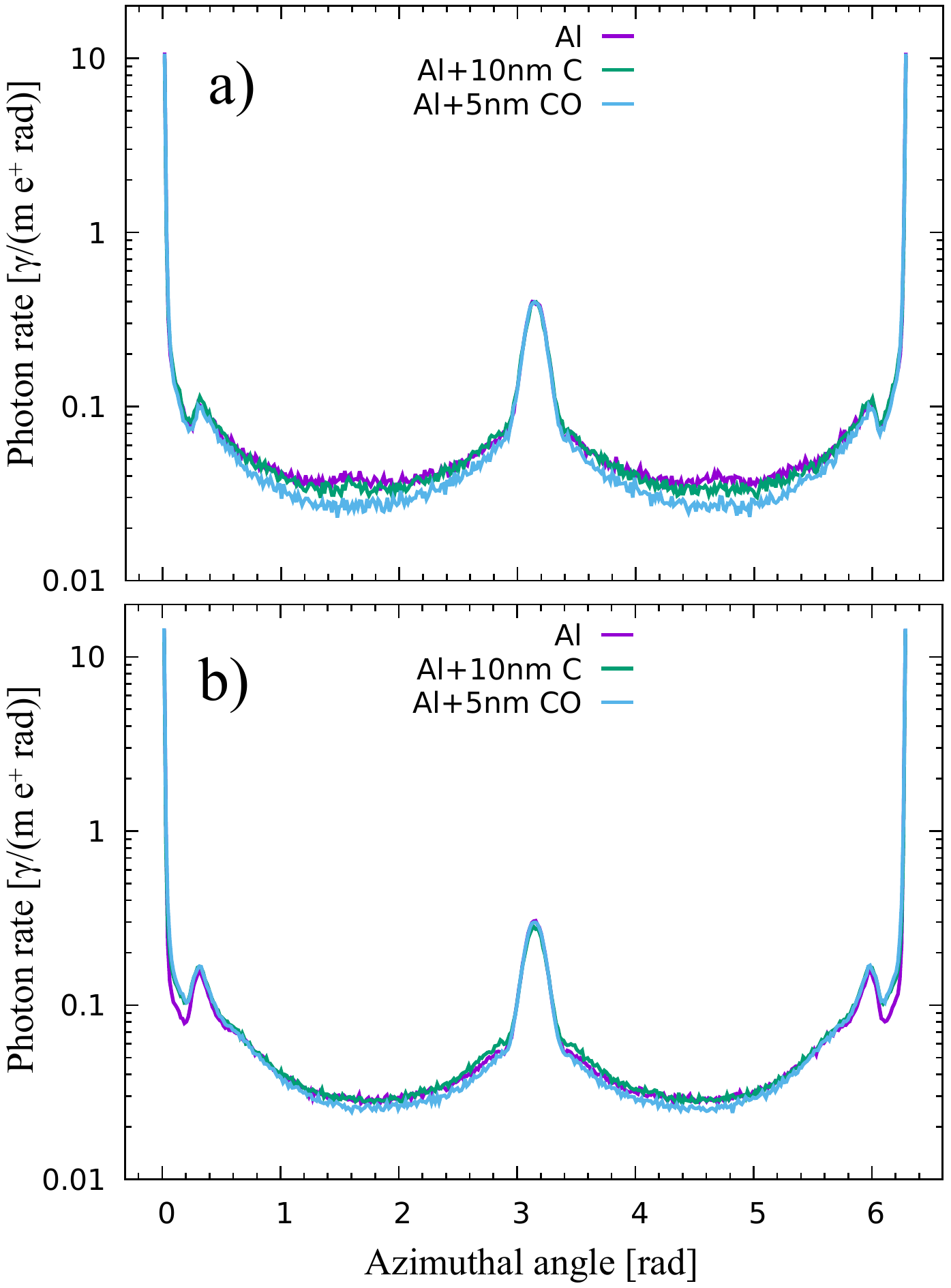}
   \caption{Azimuthal distribution of photon absorption rate averaged over a)~field-free and b)~dipole regions of the CESR ring. (Beam energy is 5.3\unit{\gev}.)
}
   \label{fig:pe_dist_ang}
\end{figure}

The photon tracking simulation thus provides the longitudinal and transverse
absorption location, and incident angle and energy on a photon-by-photon basis.
Figure~\ref{fig:pe_dist_ang} shows the absorbed photon rate in units of 
\mbox{$\text{photons}/(\text{m}\cdot\text{e}^+\cdot\text{radian})$}
as a function of transverse azimuthal absorption location, averaged separately over the a)~field-free and b)~dipole regions of the ring. The cases of a vacuum chamber material consisting of aluminum, aluminum with a \mbox{10-nm} carbon layer, or aluminum with a \mbox{5-nm} carbon-monoxide layer are compared.

\subsection{Geant4-based electron production\label{sec:sim_method:geant4}} 
The Geant4 simulation toolkit~\cite{ALLISON2016186,1742-6596-664-7-072021,AGOSTINELLI2003250}
combines theoretical calculations with measurement databases
to implement fast tracking and particle interaction algorithms for modeling tasks in experimental
particle physics, astrophysics, and medical applications, among others. An extensive bibliography
is available~\cite{GEANT4twiki:ivanchenko,GEANT4twiki:incerti},
including articles specifically on low-energy electromagnetic interactions
of photons and electrons~\cite{Apostolakis:1999bp,CIRRONE2010315,Apostolakis_2010,Chauvie2004}
and atomic de-excitation processes~\cite{Ivanchenko2011898,Guatelli2007585}.


\subsubsection{Quantum efficiency}
\begin{figure}[htbp]
\centering
\includegraphics[width=0.98\columnwidth]{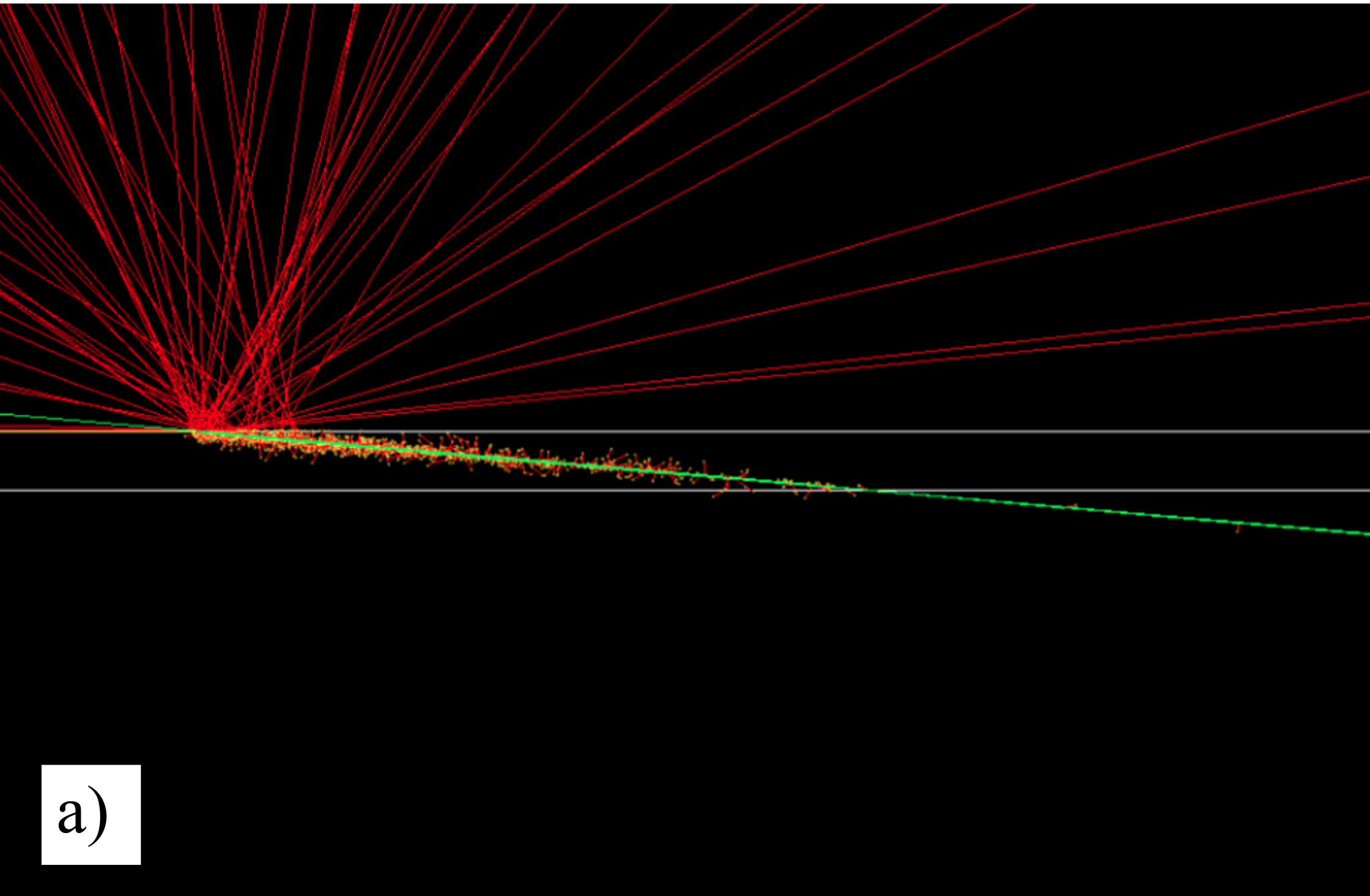}
\includegraphics[width=0.98\columnwidth]{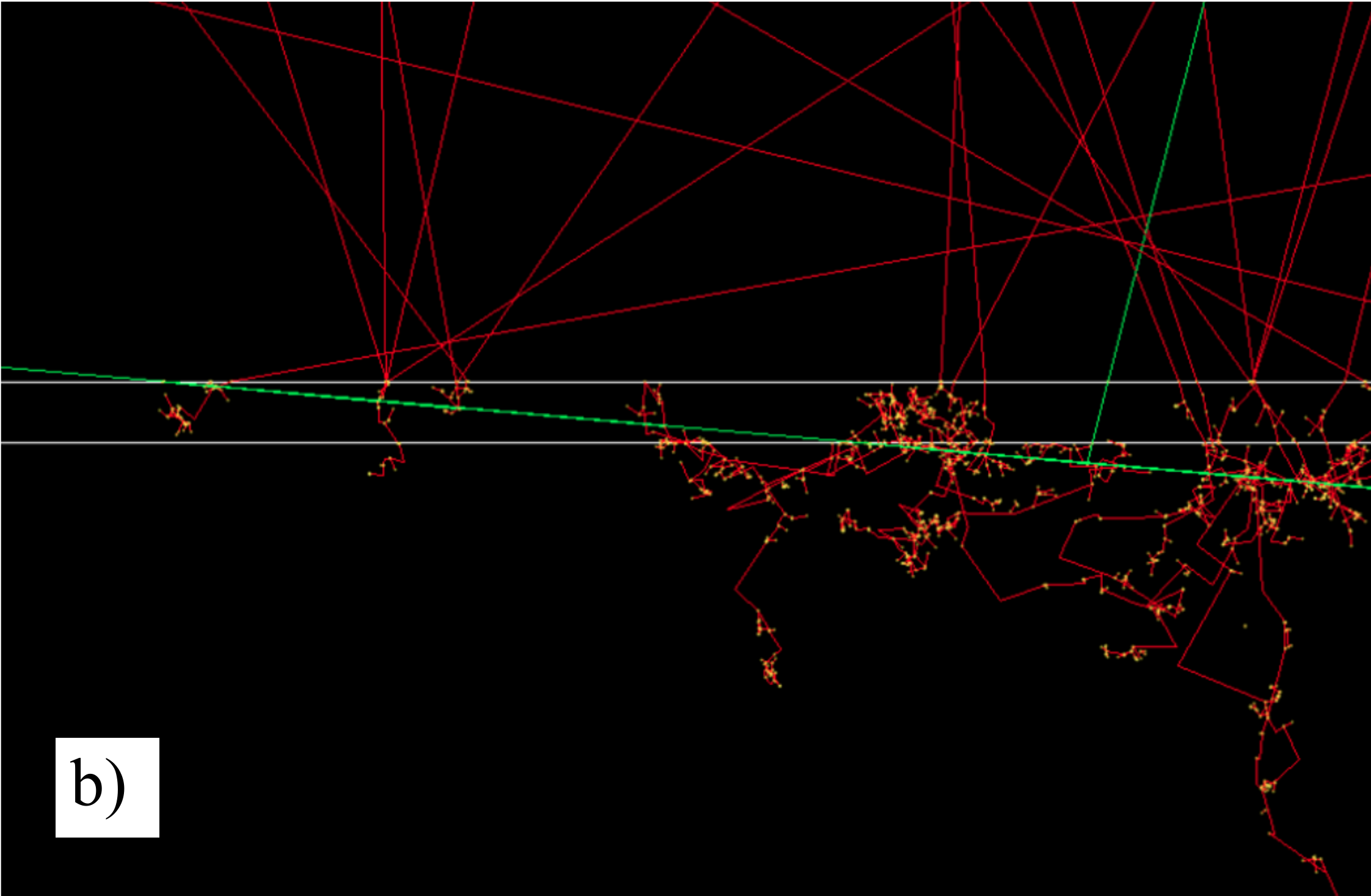}
   \caption{Tracks from incident photons (green), initially traveling left to right, and subsequently 
     generated electrons (red) in the Geant4 simulation for photon energies of a)~30\unit{\ev} and b)~2\unit{\kev}. Low-energy photons interact primarily 
     with the \mbox{5-nm} CO layer, while the higher energy photons interact in the aluminum. Electrons produced by photoeffect reach the interior
     of the vacuum chamber via re-scattering, while those produced radially symmetrically by atomic de-excitation processes can exit the wall more directly.  
}
   \label{fig:photon_events}
\end{figure}

In order to determine the azimuthal dependence of the quantum efficiency, we  
subdivide the vacuum chamber wall into 720 azimuthal bins. The grazing angle and energy
distributions of photons 
absorbed in each bin is determined by the photon tracking code. Given a sample of  photon energies and angles of incidence, 
the Geant4 code is used to generate $10^5$ photoabsorption events, determining
the rate of emitted electrons summed over the bin.
Examples of such events are shown in Fig.~\ref{fig:photon_events}. 

\begin{figure}[htbp]
\centering
\includegraphics[width=\columnwidth]{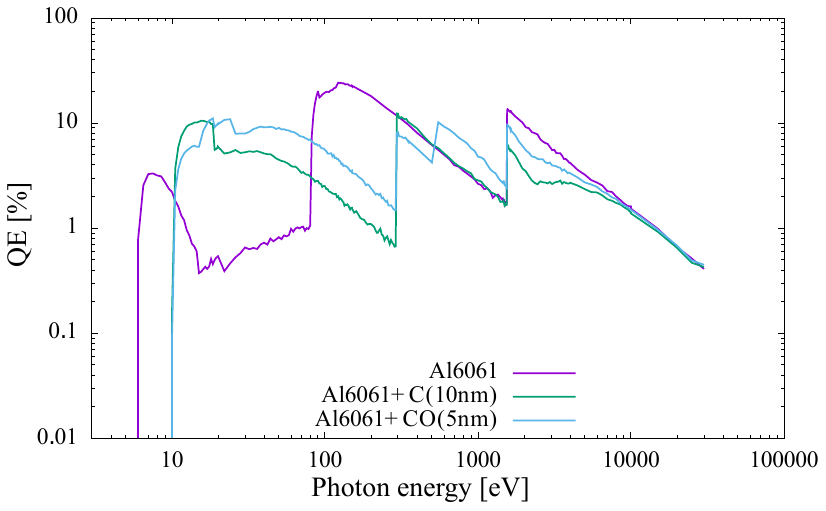}
\caption{Quantum efficiency versus photon energy for photons incident 
   at a $5^\circ$ grazing angle, for the aluminum alloy 6061, aluminum with carbon layer, 
   and aluminum with carbon monoxide layer. The quantum efficiency is sharply enhanced at photon energies 
   above various atomic shell transition energies, such as
   aluminum $\text{L}_\text{II}$ and $\text{L}_\text{III}$ (73\unit{\ev}), 
   carbon K (284\unit{\ev}), oxygen K (543\unit{\ev}), and aluminum K (1560\unit{\ev}).
   }
   \label{fig:qe_vs_e}
\end{figure}

We thus obtain a value for the electron production rate specific to the photon incident angle and energy distribution in each azimuthal bin, including (relatively rare) multi-electron production events. Figure~\ref{fig:qe_vs_e} exemplifies
the detail with which Geant4 calculates average electron production rates for various wall materials. Sharp enhancements in electron production are shown for
photon energies at the atomic shell transition energies, such as
aluminum $\text{L}_\text{II}$ and $\text{L}_\text{III}$ (73\unit{\ev}), 
carbon K (284\unit{\ev}), oxygen K (543\unit{\ev}), and aluminum K (1560\unit{\ev}).

\begin{figure}[htbp]
\centering
\includegraphics[width=\columnwidth]{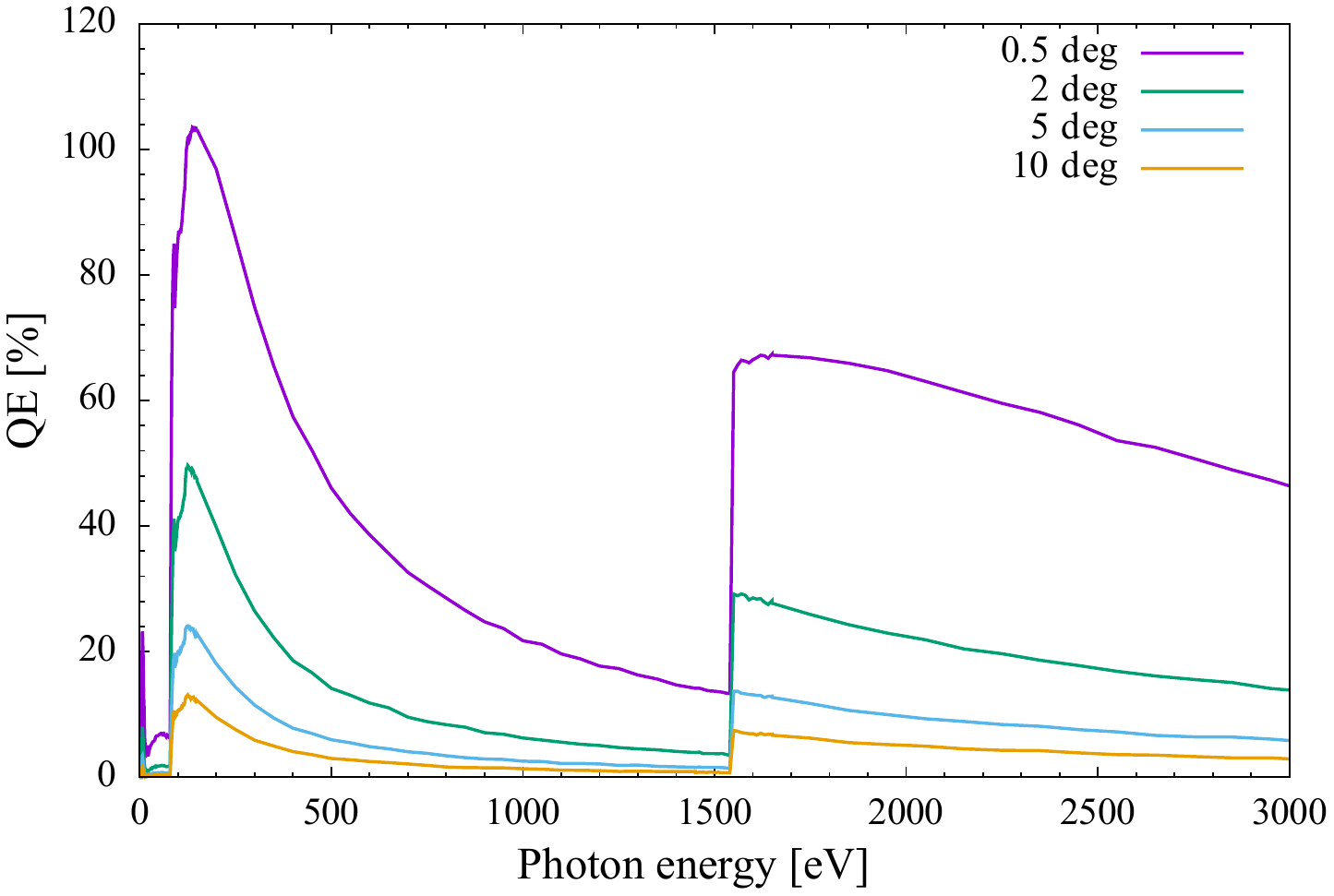}
\caption{Quantum efficiency versus photon energy for photons incident 
   at grazing angles between $0.5^\circ$ and $10^\circ$ for the aluminum alloy 6061
   as modeled in Geant4.
   }
   \label{fig:qe_degs}
\end{figure}

The strong dependence of the quantum efficiency on the incident angle of the absorbed photon 
in the Geant4 modeling is illustrated in Fig.~\ref{fig:qe_degs}, favoring more grazing angles. 
We recall that the average incident angle of the absorbed photons in the azimuthal ranges
\mbox{$|\Phi_{180}|<1.5^\circ$},
\mbox{$1.5^\circ<|\Phi_{180}|<165^\circ$} and
\mbox{$|\Phi_{180}|>165^\circ$} are $20.14^\circ$, $13.05^\circ$
and $9.66^\circ$  ($2.27^\circ$, $5.55^\circ$, and $5.77^\circ$), in the 
field-free (dipole) regions, respectively, for the case of the 5.3\unit{\gev} positron beam.

\begin{figure}[htbp]
\centering
\includegraphics[width=0.95\columnwidth]{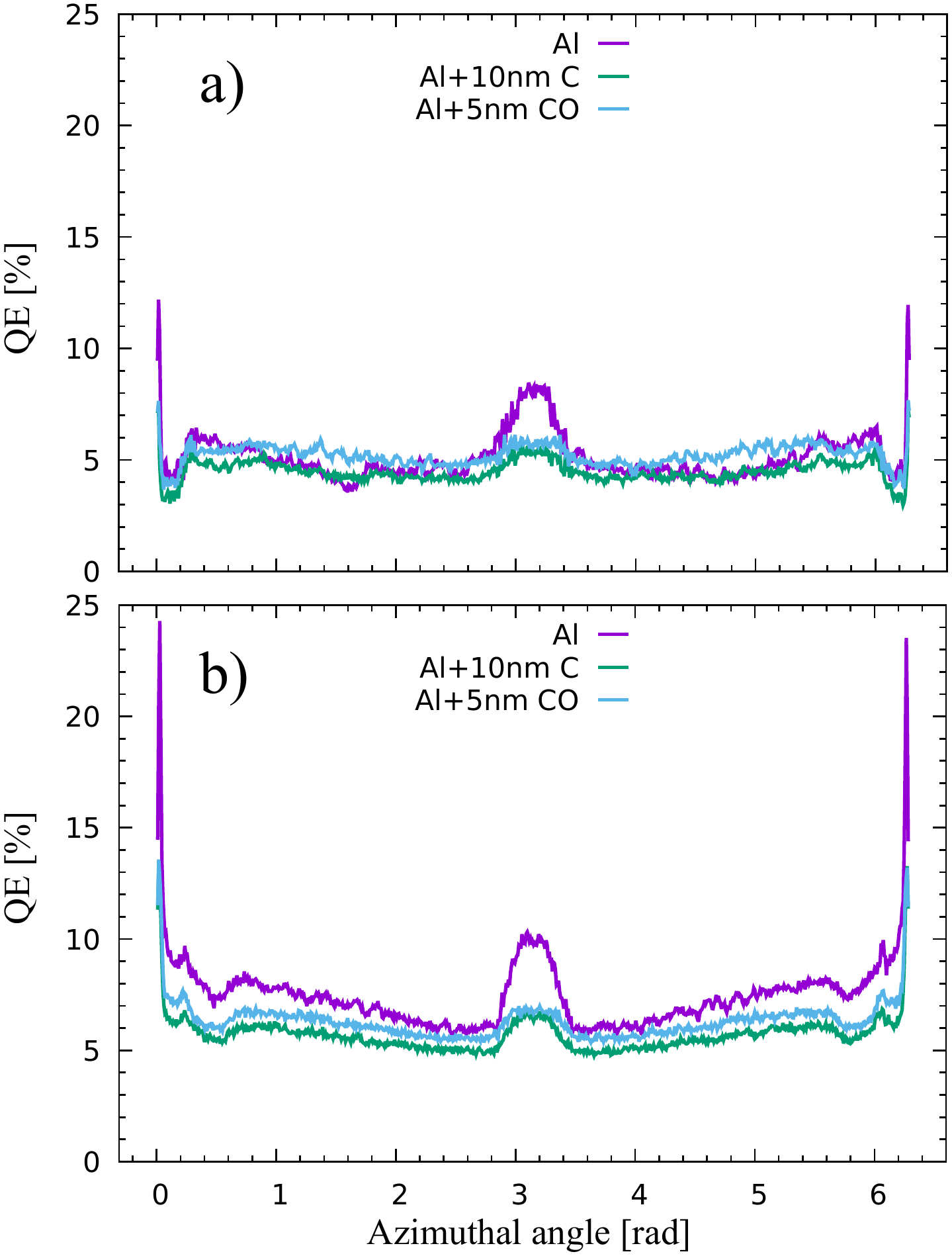}
\caption{
  Azimuthal dependence of quantum efficiency for a)~field-free regions and b)~dipole regions
of the CESR ring for aluminum and aluminum with a carbon or carbon monoxide layer. (Beam energy is 5.3\unit{\gev}.)
    }
   \label{fig:THPAF026f7}
\end{figure}

\begin{figure}[htbp]
\centering
\includegraphics[width=\columnwidth]{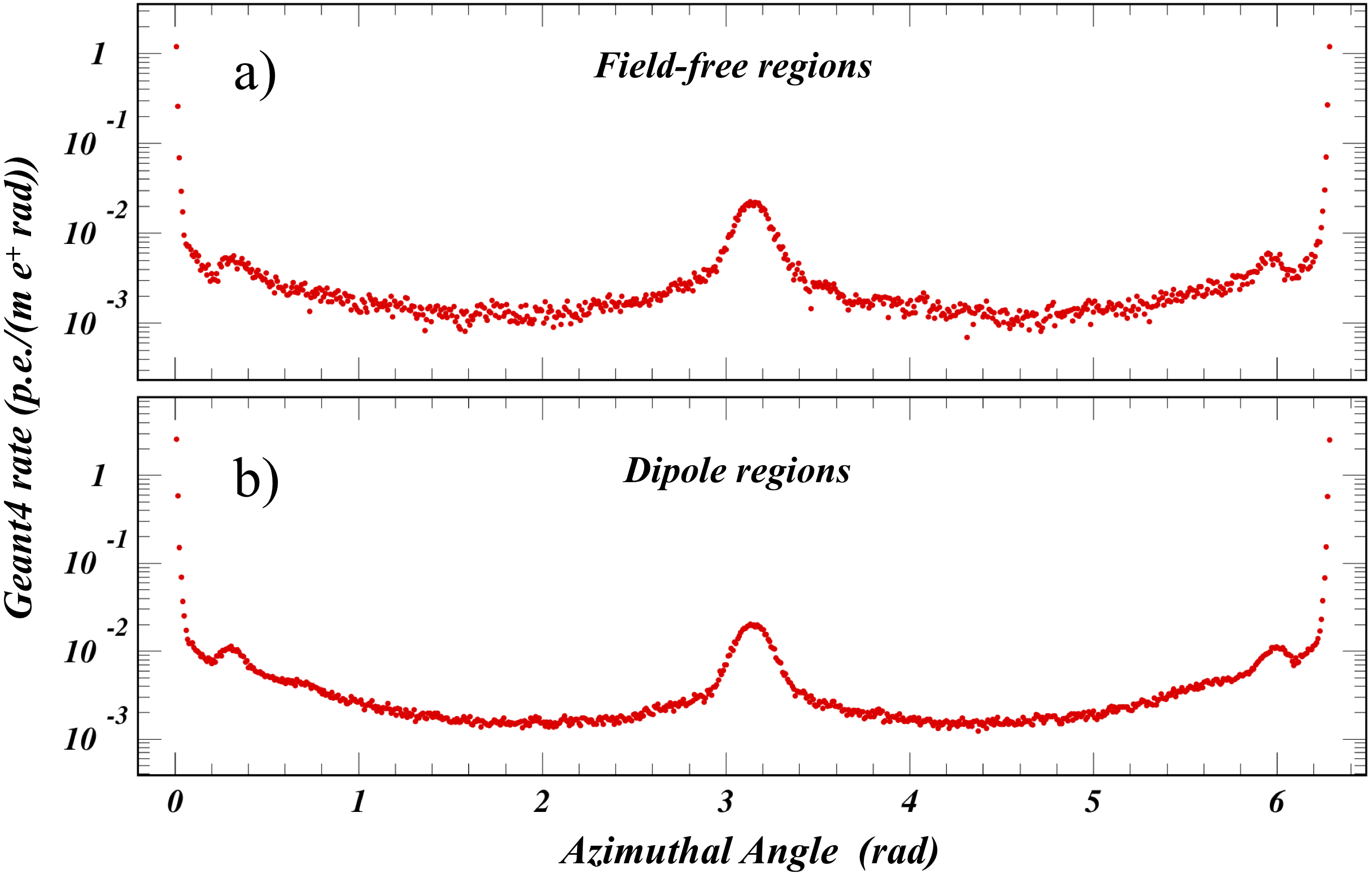}
\caption{
  Electron production rates as a function of azimuthal production location on the vacuum chamber wall for a)~field-free regions and b)~dipole regions in units of \mbox{$\text{electrons}/(\text{m}\cdot\text{e}^+\cdot\text{radian})$.} (Beam energy is 5.3\unit{\gev}.)
}
   \label{fig:pe_rates}
\end{figure}

Figure~\ref{fig:THPAF026f7} shows azimuthal distributions in average quantum efficiency obtained from the Geant4 simulations
for the 5.3\unit{\gev} positron beam.
The resulting distributions in electron production rate in the 720 azimuthal bins provided to the electron cloud buildup simulation code for the case of the aluminum chamber with the \mbox{\mbox{5-nm}} CO layer are shown in Fig.~\ref{fig:pe_rates}.
The integrated rates are 0.0454 and 0.0839 
\mbox{$\text{electrons}/(\text{m}\cdot\text{e}^+)$}
for the field-free and dipole regions, respectively.
Prior to our development work, the  photoelectron seeding for the EC buildup simulation code was characterized exclusively in terms of 
these two integrated rates and two values for effective average reflectivity around the ring~\cite{CERN:SL2002:016AP}.

\subsubsection{Photoelectron energy distributions}
\begin{figure}[htbp]
\centering
\includegraphics[width=0.95\columnwidth]{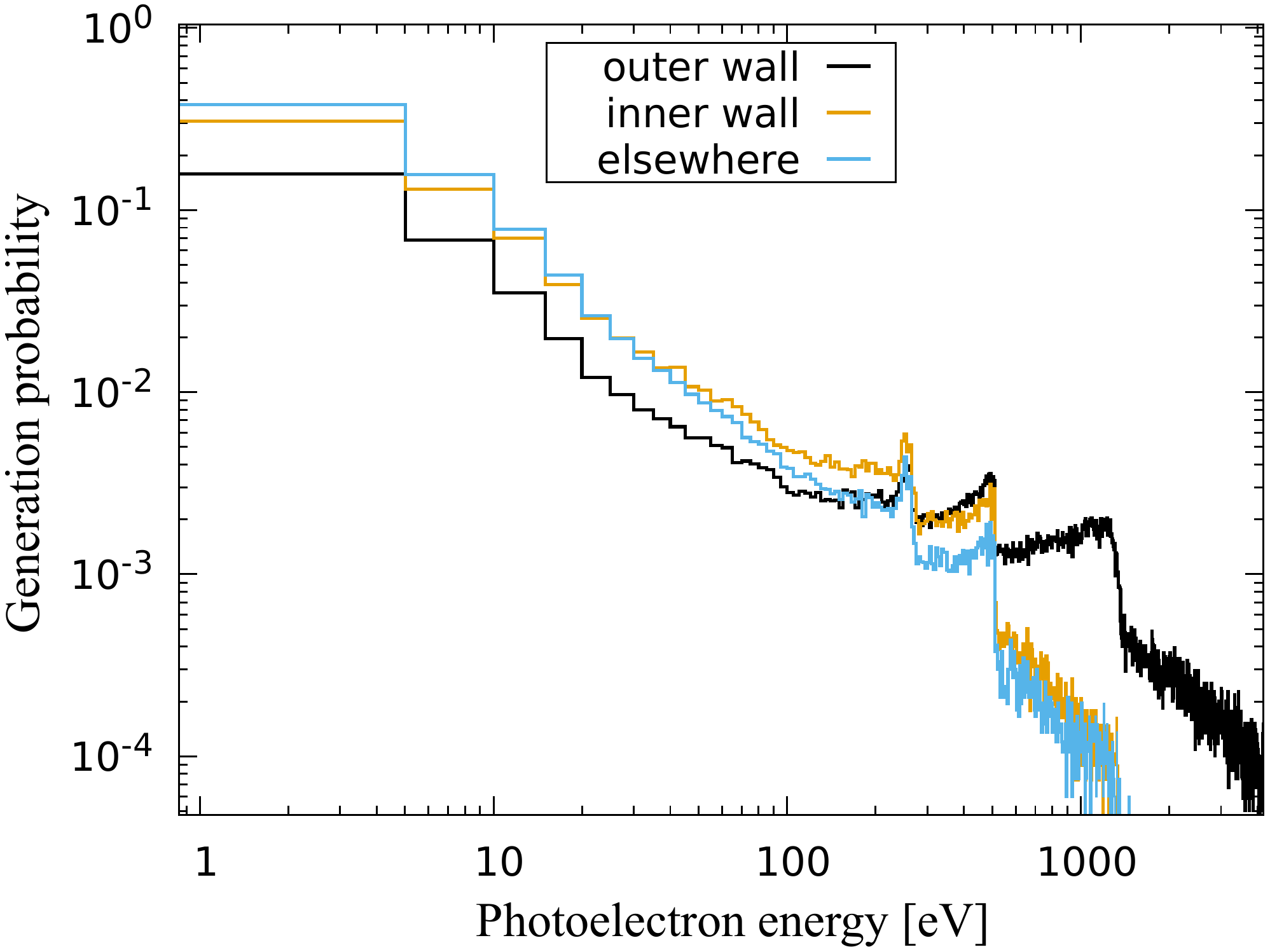}
\caption{
  The energies of photo-electrons emitted from the inner wall and elsewhere are
lower than those emitted from the outside wall. The plotted distributions are
   summed over the dipole regions. Since lower energy photons are more likely to be reflected, 
   and the inner wall and elsewhere (including top and bottom) are struck exclusively by reflected photons, the energy of the photo-electrons
   is likewise lower.
   These distributions are used as input to the electron cloud build-up simulations. (Beam energy is 5.3\unit{\gev}.)
   }
   \label{fig:pe_energy}
\end{figure}

In addition to the determination of quantum efficiencies, we obtain energy distributions of the photoelectrons in each of the three azimuthal regions
defined above, 
\mbox{$|\Phi_{180}|<1.5^\circ$}, \mbox{$1.5^\circ<|\Phi_{180}|<165^\circ$} and \mbox{$|\Phi_{180}|>165^\circ$}
by simulating $10^6$ events in each region, again with Geant4 simulations using absorbed photon data from the photon tracking code. These
distributions are shown for the CESR dipole regions in Fig.~\ref{fig:pe_energy}.  
Within each of these three angular regions, electron energy distribution is roughly independent of azimuthal angle.   
The quantum efficiency values and photoelectron energy distributions are obtained separately for
the field-free and dipole regions of the ring, resulting in a total of $1.5 \times 10^8$ simulated events 
to provide input to the electron cloud buildup simulations.

The simulation results for the photoelectron energy distributions show substantial 
high-energy tails, resulting 
in an average energy in the azimuthal ranges
\mbox{$|\Phi_{180}|<1.5^\circ$},
\mbox{$1.5^\circ<|\Phi_{180}|<165^\circ$} and
\mbox{$|\Phi_{180}|>165^\circ$} of 761\unit{\ev}, 99\unit{\ev} and 120\unit{\ev} (662\unit{\ev}, 78\unit{\ev} and 110\unit{\ev}), for the field-free (dipole) regions,
respectively. These distributions are sensitive to the atomic level thresholds satisfied by the absorbed photon energy distributions. The principal source of the high-energy electrons are atomic de-excitation processes such as the Auger effect. Both the high-energy electrons and those emitted from the wall at low energies after
multiple scattering show similar, semi-spherically symmetric exit angle distributions, so this was
used in the modeled emission angles.  
These three energy distributions, as well as the average electron production rates in $0.5^\circ$ azimuthal bins are provided separately for the field-free and dipole regions of the CESR ring as input to the electron cloud buildup calculations.
Our modeling has shown that it is important and, to an accuracy acceptable for modeling the measurement results, sufficient, to differentiate between the field-free and dipole-occupied regions, comprising 17\% and 66\% of the ring, respectively. Buildup simulations in quadrupole and other
magnetic field
environments show the contribution to the simulated tune shift values from the remaining
17\% of the ring to be at the level of a percent.
In quadrupoles, electrons are constrained to migrate in small regions
along the field lines to the poles where they are absorbed, leading to a rapid attenuation of the cloud.
Simulations indicate that the 
density of the cloud along the trajectory of the beam is small in quadrupole fields~\cite{PRSTAB18:041001}.
The large-aperture electrostatic separators and RF cavities are also excluded. 

\begin{figure}[tbp]
\centering
\includegraphics[width=0.85\columnwidth]{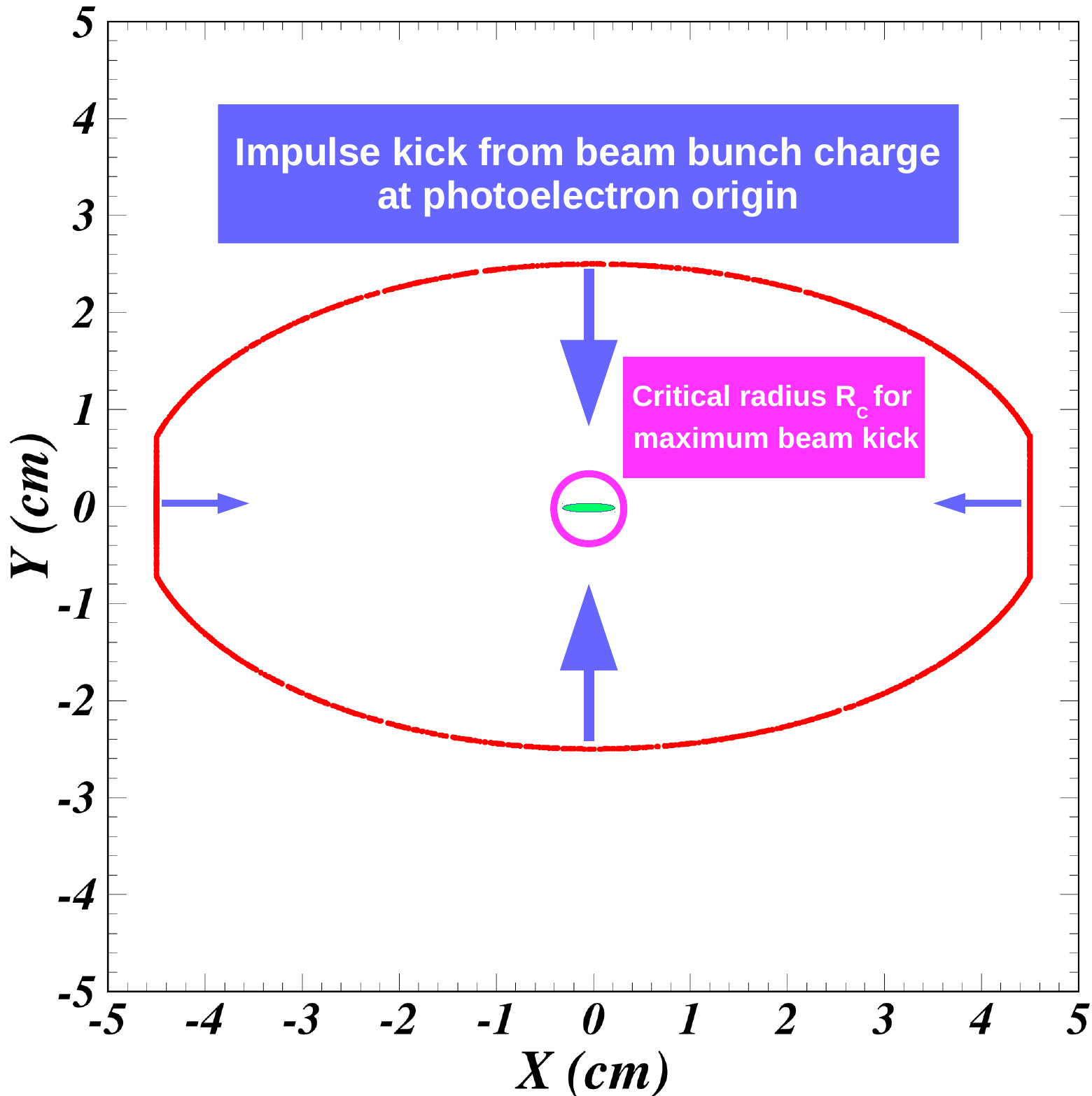}
\caption{Schematic diagram of the laterally truncated elliptical CESR vacuum chamber illustrating the 
beam kicks for an electron produced at the wall and the radius $R_{\rm C}$ at which an electron receives the maximum kick. Examples of these quantities are given in Tab.~\ref{tab:beamkick}.
       }
   \label{fig:beamkick}
\end{figure}

The energy distribution of produced electrons is of particular importance, since the modeled and measured
betatron tune shifts show a strong dependence on beam bunch population 
between $0.64 \times 10^{10}$ and $9.6 \times 10^{10}$ positrons/bunch. The associated beam kicks
for electrons produced at the wall can be comparable to the electron production energies. 
These Geant4 simulations show that the primary sources of high-energy electrons 
\mbox{($>$100\unit{\ev})} are 
atomic de-excitation processes, such as the Auger effect. The contribution of such electrons to cloud development is 
greater at lower bunch population, since their kinetic energies provide for higher subsequent SEY, 
replacing the effect of strong momentum kicks from the beam bunches. Figure~\ref{fig:beamkick} shows a 
schematic diagram of the CESR vacuum chamber illustrating the beam kick quantities in Tab.~\ref{tab:beamkick}. 
\begin{table*}[hbpt]
\centering
\caption{Parameters for the acceleration provided by a positron bunch to a cloud electron located at the
  vacuum chamber wall on the $X$ or $Y$ axes. These examples correspond to the 
  {\cesrta} measurements of betatron tune shifts as well as for the predictions for the 6.0\unit{\gev}
  upgrade of CESR~\cite{NAPAC16:TUPOB23,CLASSE:REU2018:rowan}. Kick values for the case of the field-free
  regions of the ring are shown. The total kick values are given as the
  kinetic energy
  of the electron following acceleration by the positron bunch in the impulse approximation. The direct
  and image kick values are signed according to whether they add or subtract from the total kick.
  The beam sizes shown are averages over the field-free, dipole, combined-function (DQ) magnet and
  compact undulator (CCU) regions of the ring.
        }
\label{tab:beamkick}
\linethickness{3mm}
\renewcommand{\arraystretch}{1.1} 
\begin{tabular}{l|l||cc|ccc|cc}
 
\hline
\hline
\multicolumn{2}{l||}{Beam energy ({\gev})} &  \multicolumn{2}{c|} {2.085} & \multicolumn{3}{c|} {5.289} & \multicolumn{2}{c} {6.000} \\
\hline
 &Drift (mm) & \multicolumn{2}{l|} {$0.856 \times 0.027 \times 9.2$} & \multicolumn{3}{c|} {$1.50 \times 0.142 \times 15.8$} & \multicolumn{2}{c} {$0.875 \times 0.043 \times 15.6$}\\
 &Dipole (mm) & \multicolumn{2}{l|} {$0.732 \times 0.026 \times 9.2$} & \multicolumn{3}{c|} {$1.44 \times 0.139 \times 15.8$} & \multicolumn{2}{c} {$0.889 \times 0.043 \times 15.6$}\\[-1ex]
\raisebox{1.5ex}{Beam size ${\sigma}_{\rm X}\times{\sigma}_{\rm Y}\times{\sigma}_{\rm Z}$}
 &DQ magnets (mm) & \multicolumn{2}{c|} {N/A} & \multicolumn{3}{c|} {N/A} & \multicolumn{2}{c} {$0.219 \times 0.040 \times 15.6$}\\
 &CCU undulators (mm) & \multicolumn{2}{c|} {N/A} & \multicolumn{3}{c|} {N/A} & \multicolumn{2}{c} {$0.566 \times 0.018 \times 15.6$}\\
\hline
\multicolumn{2}{l||}{Bunch population ({$10^{10}$})} & {0.64} & 1.12 & 3.25 & 6.66 & 9.5 & 3.52 & 7.11\\
\multicolumn{2}{l||}{Bunch current (mA/bunch) } & {0.4} & 0.7 & 2.0 & 4.2 & 6.0 & 2.2& 4.4\\
\specialrule{.1em}{.05em}{.05em}
\multicolumn{2}{l||}{Critical radius $R_{\rm C}$ (mm)} & {0.73} & 0.96 & 2.14 & 3.1 & {3.7} & 2.2 & 3.2\\
\multicolumn{2}{l||}{Maximum kick ({\kev})} & {1.2} & 2.5 & 3.5 & 9.0 & {14.1} & 6.1 & 14.0\\
\hline
& Direct kick (\ev) & 0.16 & 0.5 & 41.8 & 17.6 & 36 & 4.9 & 20.0\\
$X$=4.5, $Y$=0\unit{cm} & Image kick (\ev) & -0.14 & -0.44 & -41.3 & -15.6 & -32 & -4.3 & -17.7\\
& Total kick (\ev) & 0.02 & 0.06 & 0.5 & 2.0 & 4 & 0.6 & 2.3\\
\hline
& Direct kick (\ev) & 0.50 & 1.6 & 13.4 & 56 & 115& 15.8 & 64.4\\
$X$=0, $Y$=2.5\unit{cm} & Image kick (\ev) & 0.60 & 1.6 & 13.9 & 59 & 120 & 16.3 & 66.5\\
& Total kick (\ev) & 1.10 & 3.2 & 27.3 & 115 & 235 & 32.1 & 130.9\\
\hline
\hline
\end{tabular}
\renewcommand{\arraystretch}{1.0}
\end{table*}
In an impulse approximation, the beam bunch charge integrated over the bunch passage gives the momentum kick to an electron produced at the wall~\cite{LHC:ProjRep:97}. An electron generated simultaneously with the passage of the longitudinal center of the bunch, for example, receives half of this kick.
In Tab.~\ref{tab:beamkick}, we present the kick as the kinetic energy gained by the electron during the bunch passage. The elliptical shape of the vacuum chamber results in an increased (reduced) kick in the vertical (horizontal) plane from the image charges which ensure the boundary conditions at the wall. The transverse beam size determines the critical radius $R_{\rm C}$ at which a cloud electron receives the maximum kick during bunch passage.
Table~\ref{tab:beamkick} 
shows these values for the bunch populations and beam sizes for which {\cesrta} betatron tune shift measurements are available, and also for the parameters of the upgraded Cornell High Energy Synchrotron Source to be commissioned at 6\unit{\gev} in 2019~\cite{PhysRevAccelBeams.22.021602}.
The kick corresponding to wall-to-wall traversal of cloud electrons between bunch passages
depends on the bunch separation. For 14-ns bunch spacing
the kick for horizontal (vertical) wall-to-wall traversal prior to arrival of the succeeding bunch is
36\unit{\ev} (9\unit{\ev}).  Another relevant consideration in this regard is that SEY is
maximum for an electron carrying an energy of about 300\unit{\ev}.

The wide range of beam kick values causes a great variation in the cloud dynamics as a function of
bunch population as evidenced by the patterns of observed tune shifts.
The interplay between these kicks and the electron production energy distribution is an important
aspect of the cloud buildup, especially when they result in cloud electron energies on the
steeply rising slope
of the SEY curve. The effects of the photoelectron production energy distribution are particularly
pronounced at low bunch populations such as those for the 2.1\unit{\gev} data, where we have observed
changes in the modeled tune shifts of about 30\% when only low-energy ($\simeq$~5\unit{\ev})
photoelectrons are included in the simulations.

\subsection{Electron cloud buildup\label{sec:sim_method:ecloud}} 

\begin{figure}[b]
\centering
\hspace{-2mm}
\includegraphics[width=1.01\columnwidth]{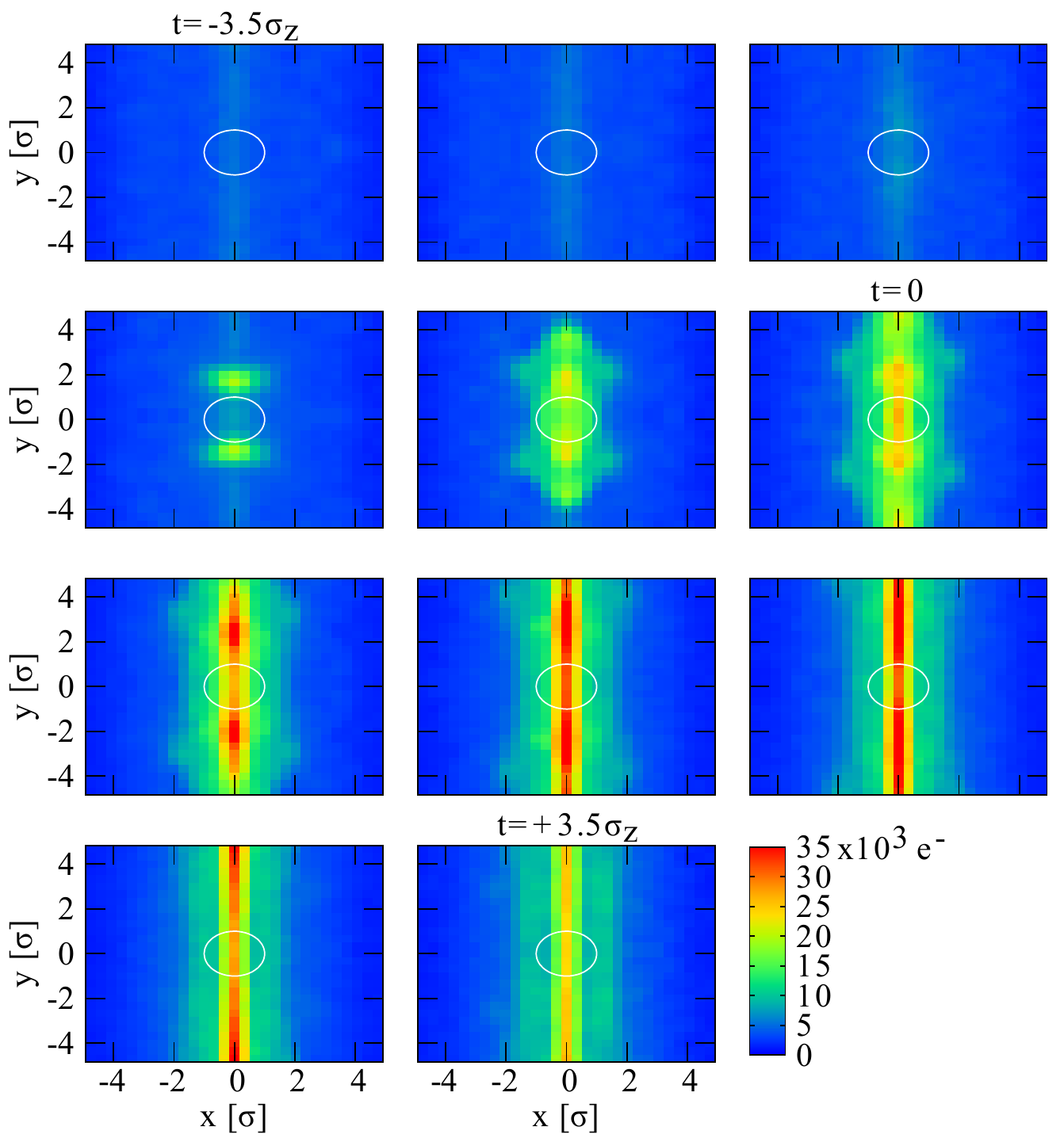}
   \caption{
   Transverse charge distributions of the electron cloud in an 800-Gauss dipole field
   during the passage of the last bunch of the 30-bunch train at 0.7\unit{mA/bunch} at 2.1\unit{\gev},
   in the central region ($\pm5\sigma$ of the beam size) for 11~time slices spanning~$\pm 3.5\sigma_z$. 
   The RMS beam size is shown as a white circle.
   Time increases from left to right, top to bottom. The time between slices is 20\unit{ps}.
   (Beam energy is 2.1\unit{\gev}.)
   }
   \label{fig:ecloud_charge}
\end{figure}

The EC buildup simulation is based on an extended version~\cite{IPAC10:TUPD024} of the ECLOUD~\cite{ICFABDNL33:14to24} code. 
The number of primary electrons created by each beam particle, along with energy and angular distribution, is input to the buildup simulation.
That information is derived from the photon tracking and electron production simulations described in the previous sections.
The buildup of the cloud is largely determined by the emission of secondary electrons from the walls of the vacuum chamber. In these simulations, the phenomenology of the
SEY physics is a parameterized Furman-Pivi model~\cite{PRSTAB5:124404}.
The SEY parameters are fit to data as described in Sec.~\ref{sec:sim_method:sey_optimization}.
Additional inputs to the buildup simulation include beam size and bunch population as given in
Tab.~\ref{tab:beamkick}. The beam sizes used in these simulations for the 2.085\unit{\gev}, 5.289\unit{\gev}, and 6.0\unit{\gev} beams are averaged over the field-free, dipole, combined-function (DQ) and compact undulator (CCU) regions of the ring.
The large ring-averaged horizontal size is dominated by dispersion.
In these simulations we clearly see the pinch effect of the beam attracting the EC (Fig.~\ref{fig:ecloud_charge}). 
We find that cloud buildup is rather insensitive to the beam size, and that using ring-averaged values per element type is a sufficient approximation. The following figures in this section characterizing the cloud buildup model use the example of the
2.1\unit{\gev} simulations.

\begin{figure*}[htbp]
\centering
\includegraphics[width=0.65\textwidth]{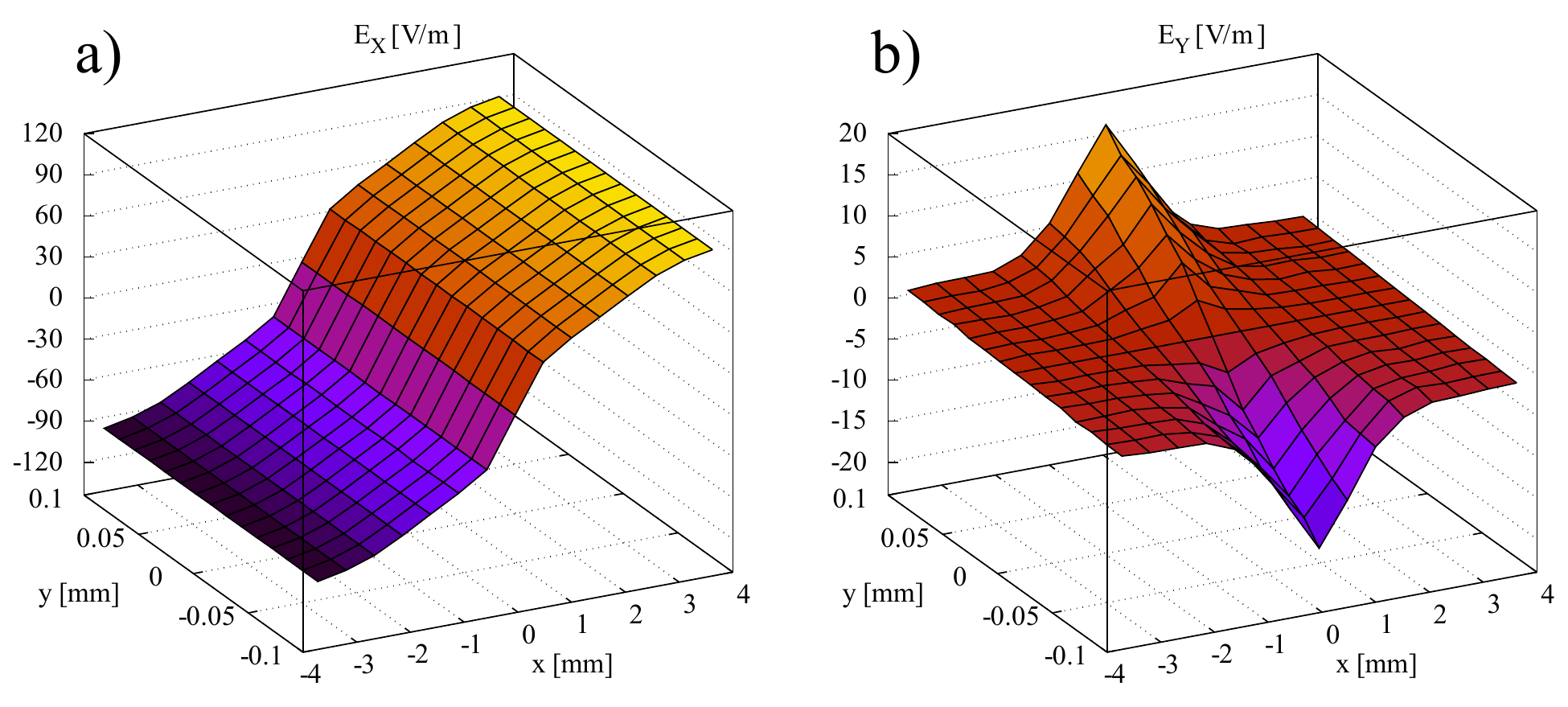}
   \caption{
   Space-charge electric field maps a)~\mbox{$E_{\rm X}(x,y)$} and b)~\mbox{$E_{\rm Y}(x,y)$} in a region of $\pm5\sigma$ of the transverse beam size 
   for the central time slice of 
   the last bunch of the 30~bunch train at 0.7\unit{mA/bunch} at 2.1\unit{\gev}, corresponding to the sixth picture in Fig.~\ref{fig:ecloud_charge}.
   (Beam energy is 2.1\unit{\gev}.)
   }
   \label{fig:efield}
\end{figure*}

Electric field maps on a $15\times 15$ grid of $\pm 5\sigma$ of the transverse beam size are obtained for 11 time slices as the bunch passes through the cloud. 
The time interval between slices is 20\unit{ps}. 
Figure~\ref{fig:efield} 
shows these field maps in a dipole for bunch number 30 in the 0.7\unit{mA/bunch} train during the central time slice.
Since only a small fraction ($\sim$$0.1\%$) of photoelectrons are within the $\pm 5\sigma$ region around the beam, it is necessary to combine the
results of many ECLOUD simulations to achieve sufficient statistical accuracy in the calculation of the electric field.

The modeled horizontal tune shift values are calculated from the cloud space-charge electric field gradients according to
\begin{equation*}
\Delta Q_x = f_{\rm rev} \frac{e}{4\pi E_{\rm beam}} \oint \beta_x \Big\langle \frac{\ud E_{\rm X}}{\ud x} \Big\rangle_{\rm beam} \ud s, 
\end{equation*}
where $f_{\rm rev}$ is the revolution frequency of 390\unit{kHz}, $e$~is the electron charge,
$E_{\rm beam}$ is the beam energy, and $\langle\ud E_{\rm X}/\ud x\rangle_{\rm beam}$ is the
electric field gradient averaged over the transverse charge distribution of the beam.
The vertical tune shifts are calculated similarly.
The positive signs of the measured horizontal and vertical tune shifts
(Figs.~\ref{fig:5gev_tuneshift_meas} and \ref{fig:2gev_tuneshift_meas}) indicate
that ${\vec\nabla\cdot \vec E} \neq 0$ and that it is cloud electrons in the path of the
positron bunch
that are largely responsible for the tune shift. 
The like-sign behavior is similar to the beam-beam tune shift in an electron-positron collider.

The integral over ring circumference is approximated as a
sum over the field gradient calculated for each element type weighted by its ring occupancy
fraction. The beta function factor is approximated as an average of the beta function over
each element type in the ring.
Table~\ref{tab:ecmodels}
\begin{table*}[hbpt]
\centering
\caption{Modeling results and parameter values used in each of the simulated tune shift calculations.
  The vacuum chamber shapes used in the EC buildup simulations are approximately elliptical with vertical side walls,
  except for the undulator chambers, which are rectangular. The numbers of photons and electrons refer to the
  total numbers generated in the simulation.
        }
\label{tab:ecmodels}
\linethickness{3mm}
\renewcommand{\arraystretch}{1.1} 
\begin{tabular}{l|cc|cc|cccc}
\hline
\hline
Beam energy (\gev) &  \multicolumn{2}{c|} {2.085} & \multicolumn{2}{c|} {5.289} & \multicolumn{4}{c} {6.000} \\
\hline
& Field-free & Dipole & Field-free & Dipole  & Field-free & Dipole & DQ magnet & CCU\\
\hline
Ring fraction (\%) &16.3 & 65.7&16.3&65.7&57.1&23.1&3.7&2.9\\
Number of photons &$1.72{\times}10^5$&$7.10{\times}10^5$&$1.56{\times}10^5$&$7.57{\times}10^5$&$7.64{\times}10^5$&$3.26{\times}10^6$&$3.37{\times}10^5$&$7.82{\times}10^4$\\
Photon absorption rate ($\gamma/({\rm m}\cdot{\rm e}^+)$) &0.378&0.370&0.728&0.876&0.833&0.973&1.655&0.3076\\
Number of electrons &$3.65{\times}10^6$&$4.54{\times}10^6$&$3.74{\times}10^6$&$4.55{\times}10^6$&$3.88{\times}10^6$&$4.58{\times}10^6$&$4.65{\times}10^6$&$4.49{\times}10^6$\\
Electron production rate (${\rm p.e.}/({\rm m}\cdot{\rm e}^+)$) &0.02137&0.03144&0.0454&0.0839&0.0603&0.0956&0.1241&0.0317\\
$\langle \beta_{\rm x} \rangle$ (m) & 16.80 & 16.50 & 18.00 & 17.00 & 14.10 & 13.10 & 1.77 & 11.07\\
$\langle \beta_{\rm y} \rangle$ (m) & 24.40 & 22.90 & 21.85& 21.70 & 18.10 & 19.60 & 15.70 & 3.49\\
\hline
\multicolumn{9}{l}{\vspace{-1.8ex}}\\
\multicolumn{9}{l}{EC buildup model input parameters:}\\
\hline
Vacuum chamber size (H$\times$V) (mm) & \multicolumn{2}{c|}{90$\times$50} & \multicolumn{2}{c|}{90$\times$50} & {90$\times$50} & {90$\times$50} & 50$\times$22 & 50$\times$4.5\\
Dipole field (T) & 0 & 0.0800 & 0 & 0.2007 & 0 & 0.2277 & 0.6509 & 1.0000\\
Quadrupole field gradient (T/m) & 0 & 0 & 0 & 0 & 0 & 0 & 8.762 & 0\\
\hline\hline
\end{tabular}
\renewcommand{\arraystretch}{1.0}
\end{table*}
shows modeling results and parameter values for each of the
tune shift calculations. For the upgraded light source operation at 6\unit{\gev},
contributions from the newly introduced DQ magnets and the CCU undulators
were included. The CCU magnetic field was modeled as a dipole field.

\begin{figure}[t]
\centering
\includegraphics[width=0.82\columnwidth]{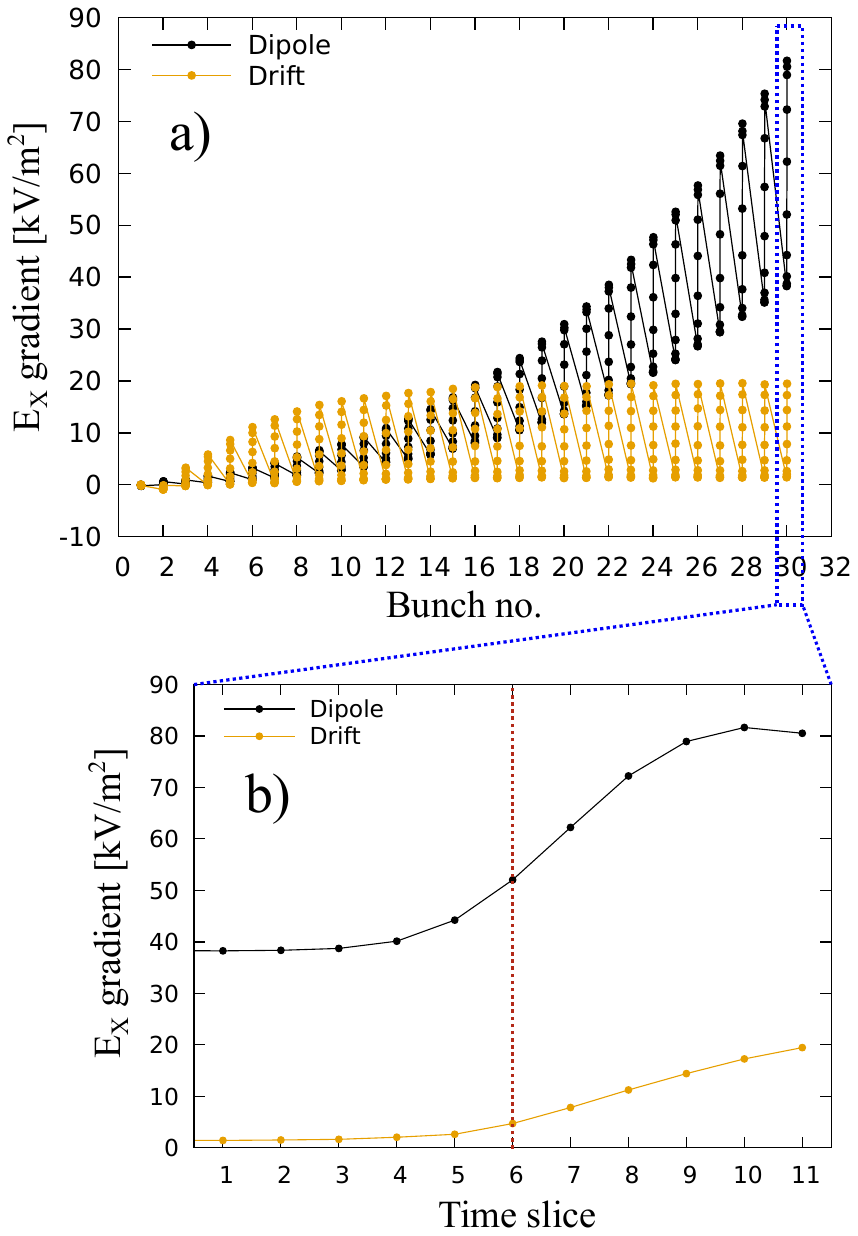}
   \caption{a)~Horizontal electron cloud 
     space-charge electric field gradients for the 11~time slices within each of 30~bunches
     for dipoles and field-free regions. b)~Electric field gradients for the 11~time slices in
     bunch~30, showing the center of the bunch at time slice~6.
   (Beam energy is 2.1\unit{\gev}.)
   }
   \label{fig:tuneshift_slices_h}
\end{figure}

\begin{figure}[htbp]
\centering
\includegraphics[width=0.82\columnwidth]{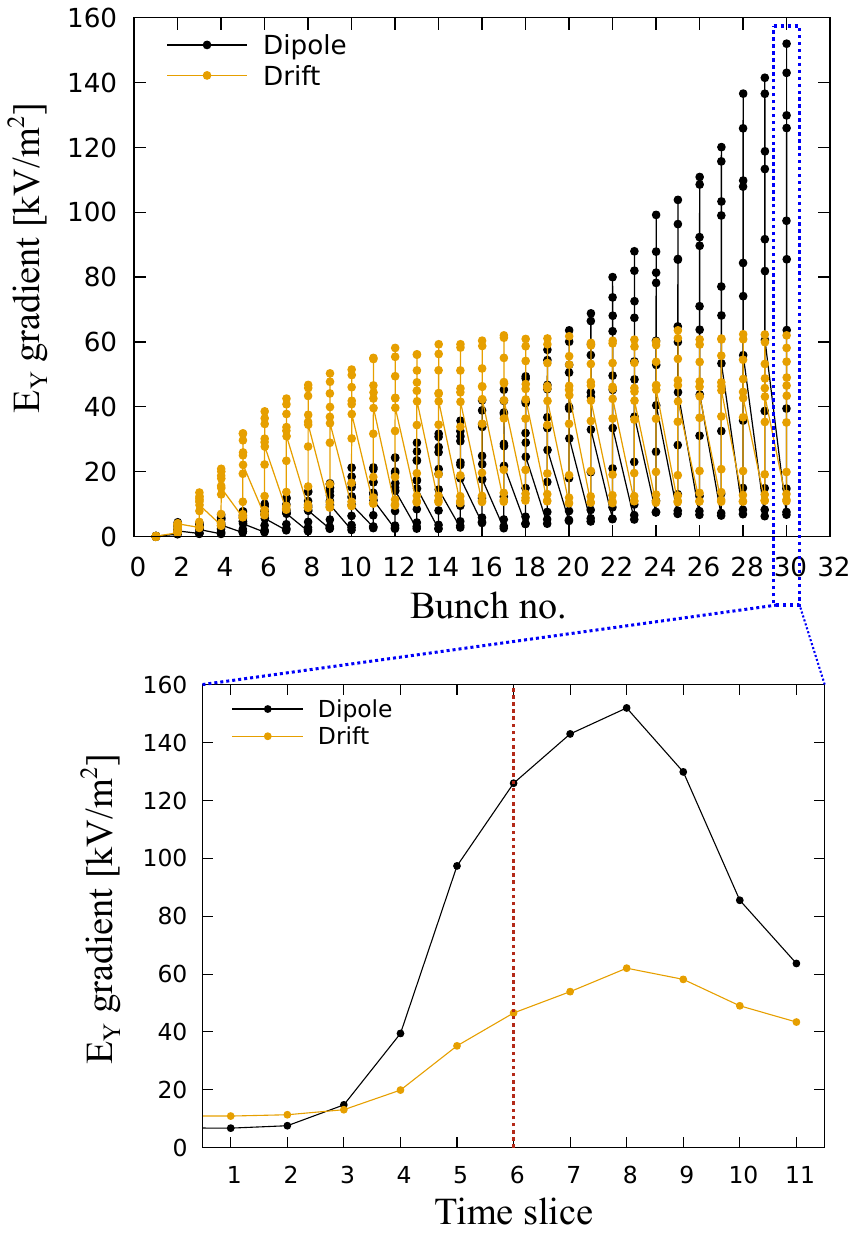}
   \caption{a)~Vertical electron cloud 
     space-charge electric field gradients for the 11~time slices within each of 30~bunches
     for dipoles and field-free regions. b)~Electric field gradients for the 11~time slices
     in bunch~30, showing the center of the bunch at time slice~6.
   (Beam energy is 2.1\unit{\gev}.)
   }
   \label{fig:tuneshift_slices_v}
\end{figure}

\begin{figure}[htbp]
\centering
\includegraphics[width=0.89\columnwidth]{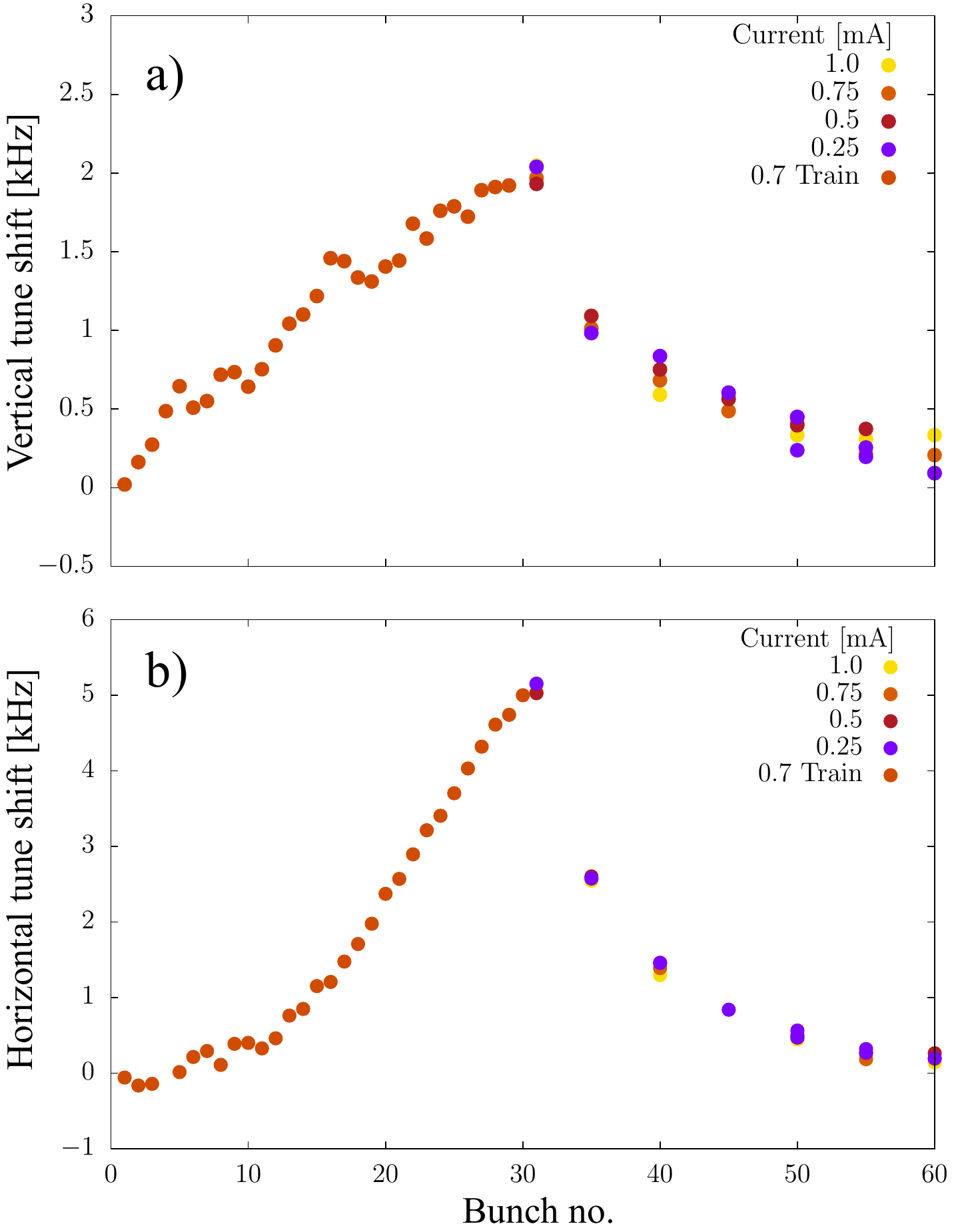}
   \caption{
   Tune shifts measured in the a)~vertical and b)~horizontal planes using the tune tracker
   for a 30-bunch train of positrons at 0.7\unit{mA/bunch} 
   ($1.12\times10^{10}$ bunch population) at 2.1\unit{\gev}, followed by a witness bunch in
   bunch positions 31--60 at currents of 0.25, 0.5, 0.75, and 1.0\unit{mA}.
   The vertical tune shift from impedance \mbox{($\simeq$$-1.0\unit{kHz/mA}$)} has been subtracted to show
   only the contribution from electron cloud.
   No dependence of the tune of the witness bunch on the witness bunch current is seen,
   showing that the pinch effect does not contribute to the tune shift.
   }
   \label{fig:witness_tunes}
\end{figure}

The pinch effect, whereby the bunch attracts the nearby cloud as it passes, can clearly be seen in
Figs.~\ref{fig:tuneshift_slices_h} and~\ref{fig:tuneshift_slices_v}
as a dramatic increase in
the modeled electric field gradients during the bunch passage.
Figure~\ref{fig:witness_tunes} shows the measured
tune shift in each of the 30~bunches in the train as well as for witness
bunches positioned one at a time with some delay beyond the end of the train.
Unlike the measured tune shifts along the train, which are referenced
to that of the first bunch in the train and where
the bunch populations are equal at a level of better than~1\%,
the observed witness bunch tune shifts require a correction for the
ring impedance contribution to the coherent tune shift, which has been measured
to be about $-1\unit{kHz/mA}$~\cite{IPAC15:MOPMA056}.
The cloud-induced
tune shifts of the witness bunches are observed to be independent of
witness bunch current, whereas the pinch by definition is not.
These measurements clearly show that the pinch effect does not contribute
to the tune shift. For this reason, the space-charge electric field gradients
immediately prior to the bunch arrival are used when calculating the modeled
tune shifts.

\begin{figure}[htbp]
\centering
\includegraphics[width=0.87\columnwidth]{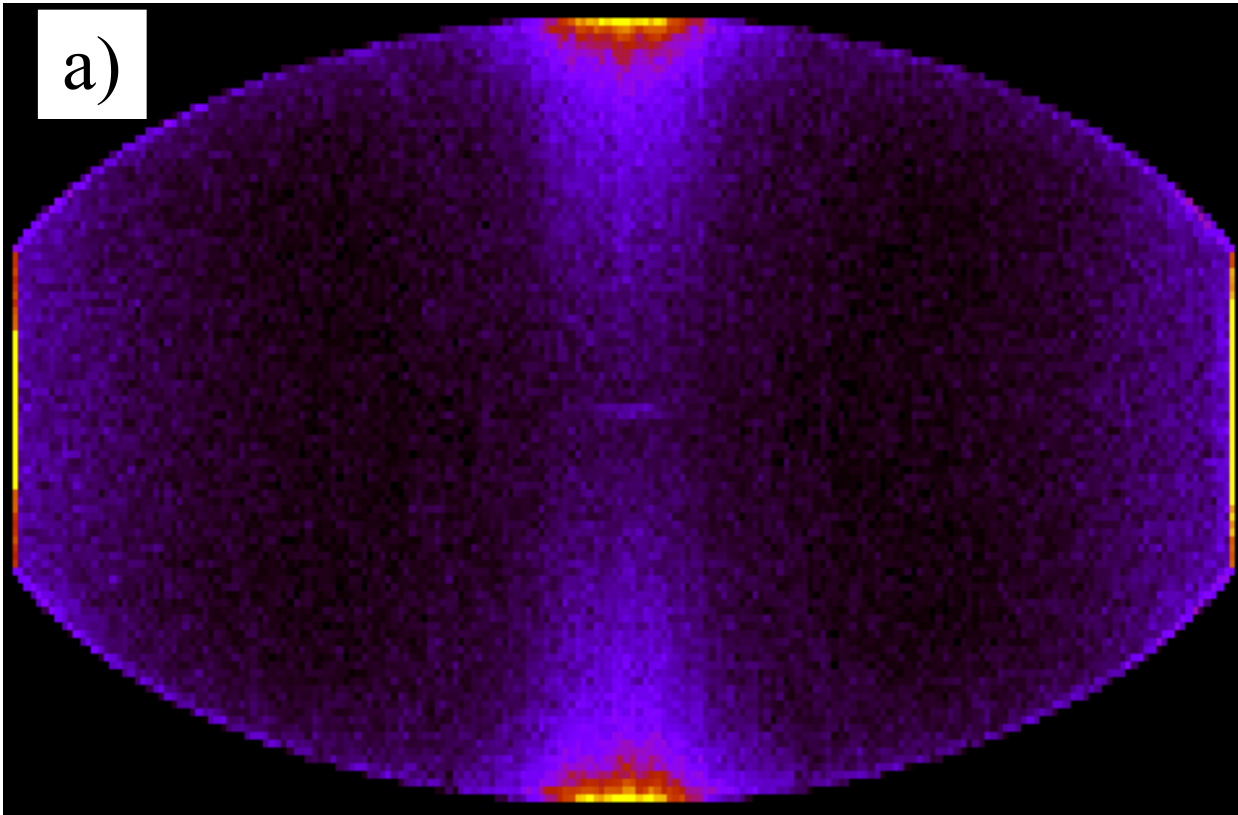}
\includegraphics[width=0.87\columnwidth]{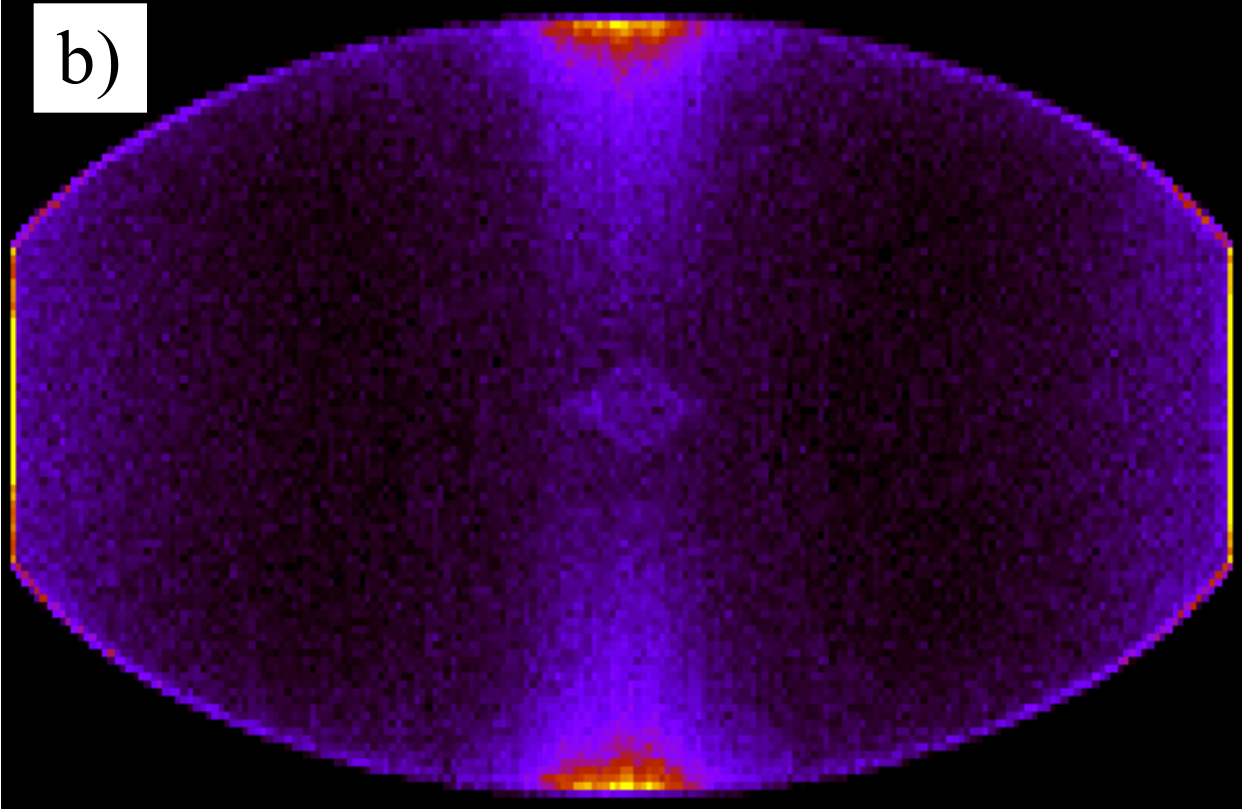}
\caption{Simulated electron cloud density during the a)~third and b)~sixth of
  11~time slices during of the passage of bunch~15, which has been offset from the
    centered bunch train by 1\unit{mm} horizontally to simulate the effect of kicking a single bunch when measuring its tune. The ``pinched'' cloud is found to be centered on the offset 
    bunch position.
    The short bunch length (16\unit{mm}) bunch  results in
    little effect
    on the larger built-up cloud.
    The simulated bunch current is 2\unit{mA/bunch}. 
    At higher currents, the vertical band widens (4\unit{mA/bunch})
    and splits into two (6\unit{mA/bunch}).
   (Beam energy is 5.3\unit{\gev}.)
   }
   \label{fig:ecloud_snapshot}
\end{figure}

Simulations for a bunch offset relative to the train are consistent with this
measurement result.
For an offset bunch (the one being excited) in an on-axis train,
the pinched cloud is
centered on the offset bunch, even in the presence of a dipole field,
as shown in Fig.~\ref{fig:ecloud_snapshot}. In the absence of any such
beam/cloud offset,
the pinch does not contribute any coherent kick to the bunch.
%
%

\subsection{SEY parameter determination\label{sec:sim_method:sey_optimization}} 
\begin{figure}[htbp]
\setlength{\unitlength}{0.889\columnwidth}
\centering
\begin{picture}(0.95,0.93)(0,0)
\put(0.05,-0.005){\includegraphics[width=0.8\columnwidth]{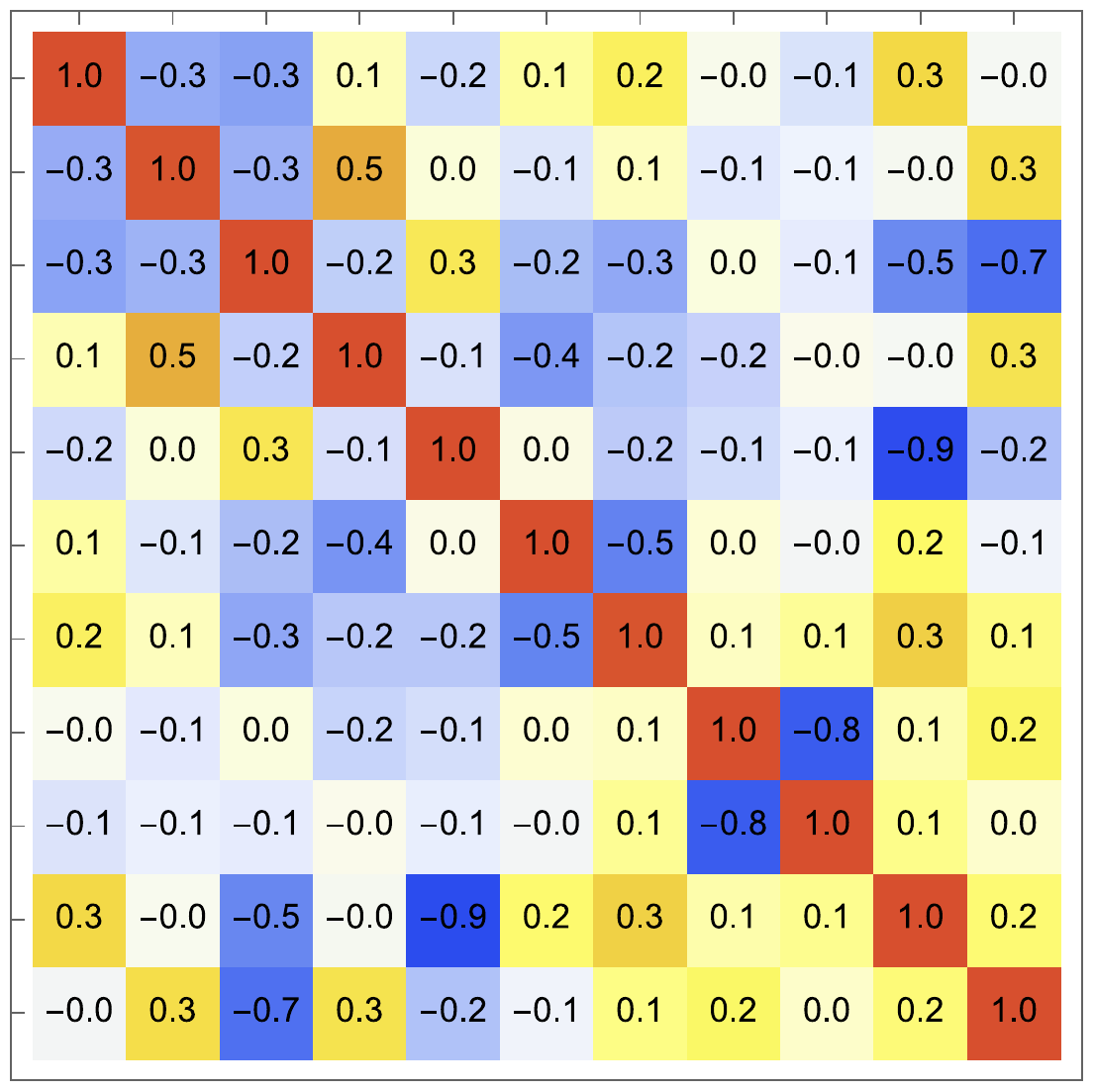}}
\put(0.09,0.9){$\hat E_{\rm ts}$}
\put(0.19,0.9){$s$}
\put(0.24,0.9){$P_{1,{\rm r}}$}
\put(0.33,0.9){$\hat\delta_{\rm ts}$}
\put(0.41,0.9){$t_1$}
\put(0.49,0.9){$t_2$}
\put(0.57,0.9){$t_3$}
\put(0.65,0.9){$t_4$}
\put(0.73,0.9){$\epsilon$}
\put(0.80,0.9){$p$}
\put(0.86,0.9){$\hat P_{1,{\rm e}}$}
\put(0.,0.82){\makebox[0.04\unitlength][r]{$\hat E_{\rm ts}$}}
\put(0.,0.75){\makebox[0.04\unitlength][r]{$s$}}
\put(0.,0.67){\makebox[0.04\unitlength][r]{$P_{1,{\rm r}}$}}
\put(0.,0.58){\makebox[0.04\unitlength][r]{$\hat\delta_{\rm ts}$}}
\put(0.,0.51){\makebox[0.04\unitlength][r]{$t_1$}}
\put(0.,0.435){\makebox[0.04\unitlength][r]{$t_2$}}
\put(0.,0.36){\makebox[0.04\unitlength][r]{$t_3$}}
\put(0.,0.285){\makebox[0.04\unitlength][r]{$t_4$}}
\put(0.,0.21){\makebox[0.04\unitlength][r]{$\epsilon$}}
\put(0.,0.135){\makebox[0.04\unitlength][r]{$p$}}
\put(0.,0.05){\makebox[0.04\unitlength][r]{$\hat P_{1,{\rm e}}$}}

\end{picture}
\caption{Correlation matrix for the optimized SEY parameters.
  Of particular note is that the rediffused yield parameter $P_{1,{\rm r}}$
  is found to be significantly higher than the initial value, while the
  elastic yield $\hat P_{1,{\rm e}}$
  is found to be lower, and these two parameters
  are highly anti-correlated.
        }
\label{fig:correlations}
\end{figure}

Secondary-electron yield
depends on a number of factors, such as incident electron energy and angle, and chamber wall material
and coatings (see list below). 
Tune shifts from simulation are found to depend strongly on these details of the SEY
The effects of the SEY parameters on the tune shifts can be highly correlated.
Direct SEY measurements can provide a good starting point, but a comprehensive experimental determination of
all the SEY parameters has yet to be obtained.
Furthermore, the ring-wide averaged SEY may be different than an external measurement
of a vacuum chamber sample.
To improve agreement between the model and the tune shift measurements, we use an optimizer to fit
the model tune shift values to the measurements, varying a selection of 11 SEY parameters.
Measurements of the peak SEY for aluminum show a rapid beam-processing-induced reduction of the
peak SEY to a value near 1.8~\cite{JVSTA23:1610to1618,LCC:0153},
comparable to the values measured for copper~\cite{EPAC00:THXF102}.
We therefore use the SEY parameters determined for copper in Ref.~\cite{PRSTAB5:124404}
as a starting point, rather than the other example given, stainless steel, for which the SEY components are
rather dissimilar.
At each iteration, the EC buildup simulations are run in parallel with the current best SEY
parameters, and each parameter increased and decreased by an adaptive increment. The tune shifts from these simulations
are obtained, and the Jacobian is calculated and provided to the optimizer.
The optimized input parameters are, in the notation of Ref.~\cite{PRSTAB5:124404},
\begin{itemize}
\setlength\itemsep{-0.1em}
\item $\hat E_{\rm ts}$: 
incident electron energy at which the true-secondary yield is maximum for perpendicular incidence,
\item $s$: true-SEY energy dependence parameter, with \mbox{$\delta_{\rm ts}(\theta_{\rm e},E_0) = \delta_{\rm ts}(\theta_{\rm e}) sx/(s-1+x^s)$}, where \mbox{$x=E_0/E_{\rm ts}(\theta_{\rm e})$}, $E_0$ being the incident electron energy,
\item $P_{1,{\rm r}}(\infty)$: rediffused SEY at high incident electron energy,
\item $\hat\delta_{\rm ts}$: true SEY at perpendicular incidence,
\item $t_1$ and $t_2$: amplitude of the cosine dependence and power of the cosine in the true SEY: \mbox{$\delta_{\rm ts}(\theta_{\rm e})=\hat\delta_{\rm ts}[1+t_1(1-\cos^{t_2}{\theta_{\rm e}}$)]}, where \mbox{$\theta_{\rm e} = 0^\circ$} for perpendicular electron incidence,
\item $t_3$ and $t_4$: amplitude of the cosine dependence and power of the cosine in true-SEY peak energy: \mbox{$E_{\rm ts}(\theta_{\rm e})=\hat E_{\rm ts}[1+t_3(1-\cos^{t_4}{\theta_{\rm e}}$)]},
\item $\hat P_{1,{\rm e}}$: elastic yield in the low-energy limit, and
\item $\epsilon$ and $p$: parameters for the energy distribution of the secondary electrons: 
\[
\frac{{\rm d} N}{{\rm d} E_{\rm sec}} (E_{\rm sec})
\propto
\begin{dcases}
\frac{{(E_{\rm sec}/\epsilon)}^{p-1} e^{-E_{\rm sec}/\epsilon}}{\epsilon} & 
    \text{for $E_{\rm sec}\leq 5\epsilon$} \\
0 & \text{for $E_{\rm sec}>5\epsilon$}
\end{dcases}
.
\]
\end{itemize}
The fits are performed simultaneously over all tune shift data at 2.1 and
5.3\unit{\gev}
shown in Figs.~\ref{fig:5gev_tuneshift_meas} and \ref{fig:2gev_tuneshift_meas}.
Table~\ref{tab:seypar} compares the optimized parameters to the initial values.
\begin{table}[htbp]
\centering
\caption{Initial and optimized SEY parameters,
  including the sensitivity
  to each parameter given in the form of an uncertainty
  calculated from the Jacobian.
        }
\label{tab:seypar}
\linethickness{3mm}
\renewcommand{\arraystretch}{1.1} 
\begin{tabular}{lccc}
\hline
\hline
& Initial & Optimized & Uncertainty \\
\hline
$\hat E_{\rm ts}$ (\ev) & 277 & 260 & 10 \\
$s$ & 1.54 & 1.58 & 0.05\\
$P_{1,{\rm r}}(\infty)$ & 0.2 & 0.39 & 0.05\\
$\hat\delta_{\rm ts}$ & 1.88 & 1.53 & 0.04\\
$t_1$ & 0.66 & 0.99 & 0.2\\
$t_2$ & 0.8 & 1.5 & 0.4\\
$t_3$ & 0.70 & 0.77 & 0.50\\
$t_4$ & 1.0 & 1.2 & 1.0\\
$\epsilon$ (\ev) & 1.8 & 3.6 & 0.4\\
$p$ & 1.0 & 0.8 & 0.2\\
$\hat P_{1,{\rm e}}$ & 0.5 & 0.07 & 0.02\\
\hline
\hline
\end{tabular}
\renewcommand{\arraystretch}{1.0}
\end{table}
The range and accuracy of the tune shift measurements provides high sensitivity
to a number of these parameters. For example, the true SEY $\hat\delta_{\rm ts}$ is determined with an accuracy of better than 3\%. On the other
hand, the
tune shifts are relatively insensitive to the peak energy angular dependence parameters $t_3$ and $t_4$. Those parameters are poorly constrained.
Correlations also limit the predictive power of the fit. The correlation matrix for the optimized SEY parameters is shown in Fig.~\ref{fig:correlations}. Note that
that
$p$ and $t_1$, parameters that characterize the angular dependence and energy distribution of secondaries are highly anti-correlated. 
The
fitted
value for the rediffused component of
the SEY ($P_{1,r}$) is found to be significantly higher than the value
obtained in Ref.~\cite{PRSTAB5:124404}; however, in view of the high degree of anti-correlation with the elastic yield value ($\hat P_{1,e}$)
 (see Fig.~\ref{fig:correlations}), which is found to be low, the uncertainty is large.
%

\section{Simulation Results\label{sec:sim_result}}

\begin{figure}[htbp]
\centering
\includegraphics[width=\columnwidth]{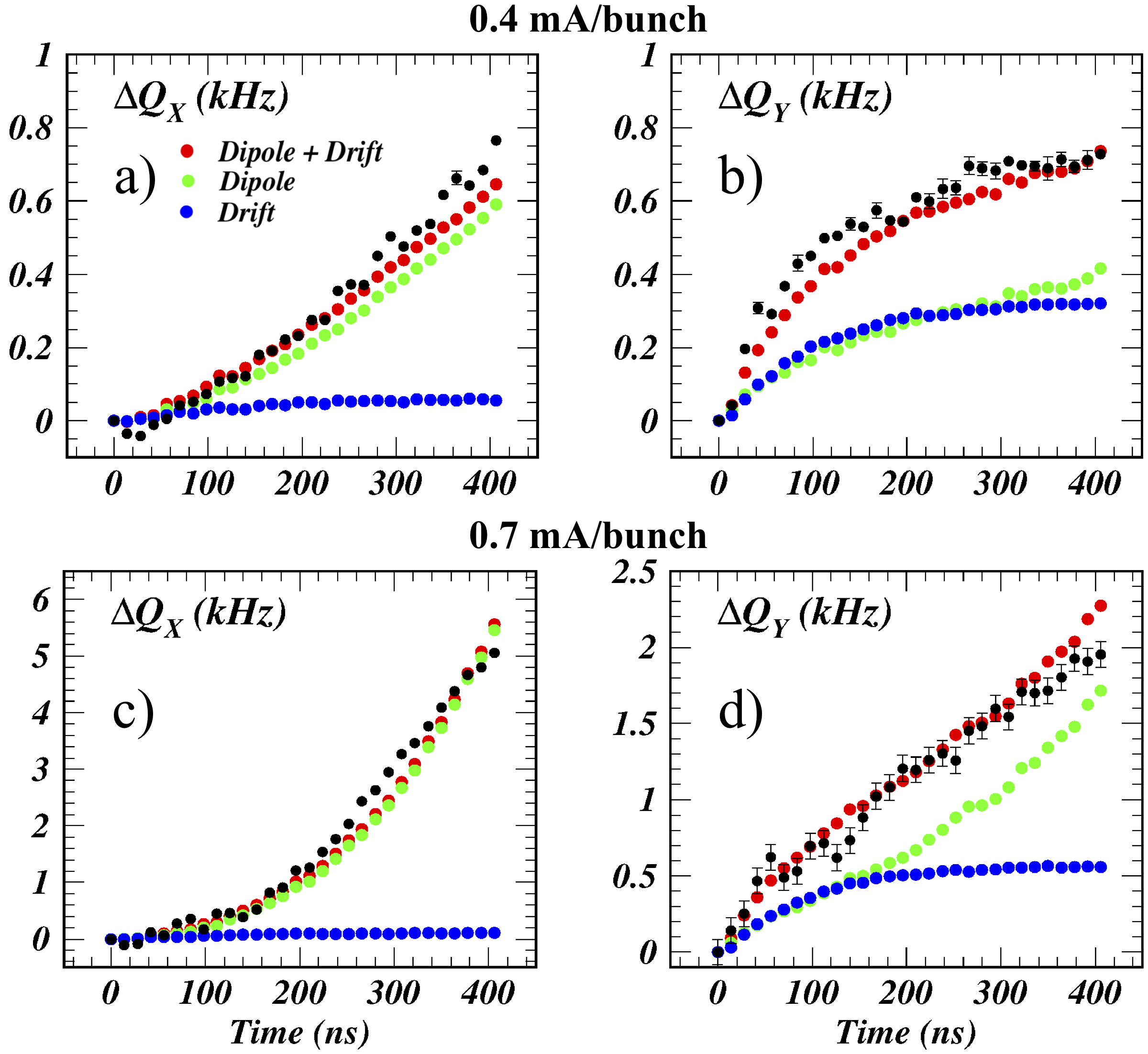}
   \caption{
     Comparison of the measured (black points) and modeled tune shift values
     for the 2.1\unit{\gev}, 30-bunch train of positrons.
     The top row shows the tune shift values in the
     a)~horizontal and b)~vertical planes for a bunch population
     of  $0.64\times10^{10}$ (0.4\unit{mA/bunch}).
     The bottom row shows the tune shift values in the
     c)~horizontal and d)~vertical planes for a bunch population
     of  $1.12\times10^{10}$ (0.7\unit{mA/bunch}).
     Contributions from the field free regions of the ring are shown in
     blue; those from the dipole regions are shown in green. The sum of the two
     contributions is shown in red.
     Bunches are spaced 14\unit{ns} apart.
   }
   \label{fig:2gev_tuneshift_sim}
\end{figure}

\begin{figure}[htbp]
\centering
\includegraphics[width=\columnwidth]{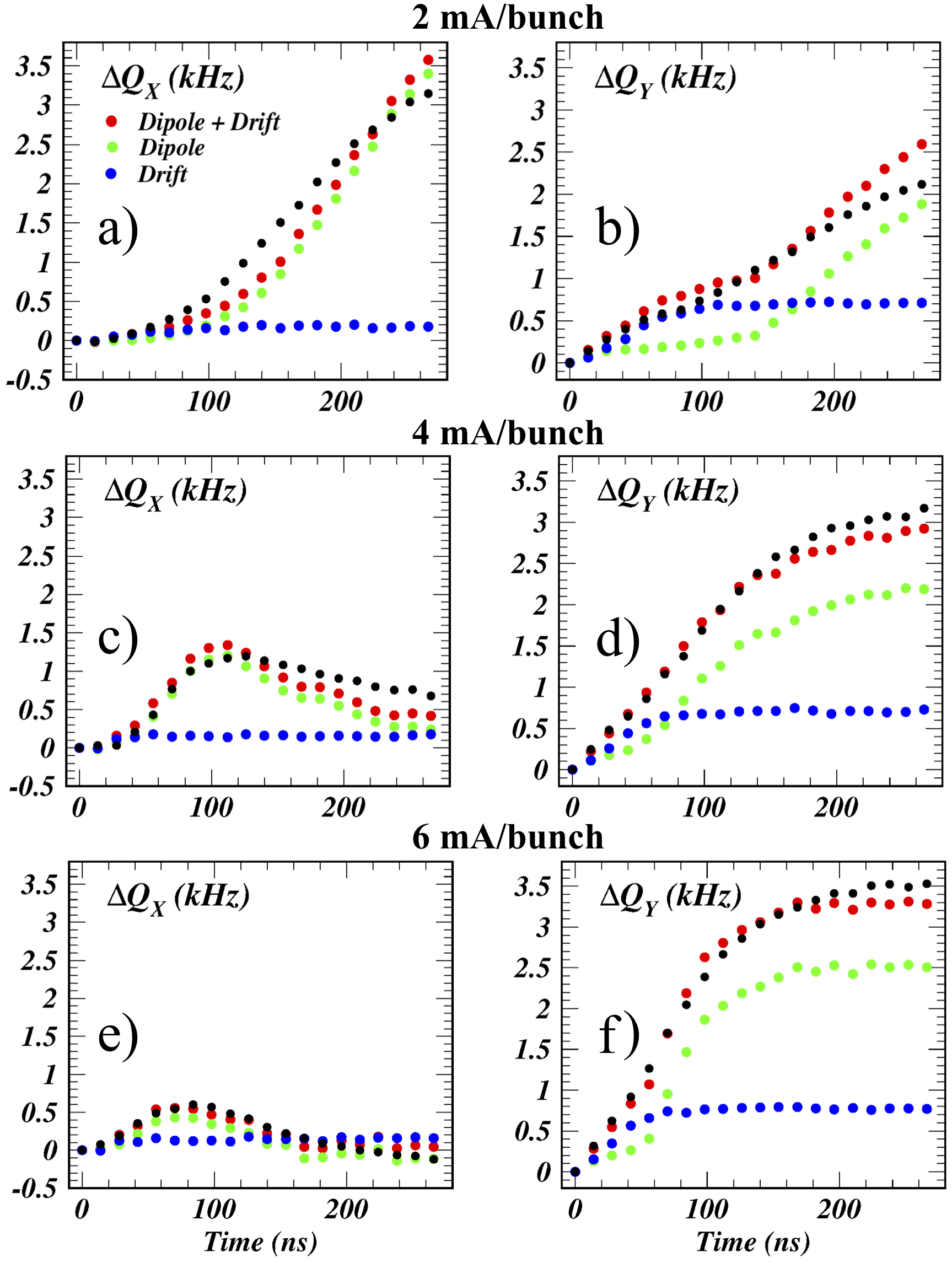}
   \caption{
     Comparison of the measured (black points) and modeled tune shift values
     for the 5.3\unit{\gev}, 20-bunch train of positrons.
     The top row shows the tune shift values in the
     a)~horizontal and b)~vertical planes for a bunch population
     of  $3.2\times10^{10}$ (2\unit{mA/bunch}).
     The middle row shows the tune shift values in the
     c)~horizontal and d)~vertical planes for a bunch population
     of  $6.4\times10^{10}$ (4\unit{mA/bunch}).
     The bottom row shows the tune shift values in the
     e)~horizontal and f)~vertical planes for a bunch population
     of  $9.6\times10^{10}$ (6\unit{mA/bunch}).
     Bunches are spaced 14\unit{ns} apart.
   }
   \label{fig:5gev_tuneshift_sim}
\end{figure}

The comparison of modeled
(using the optimized SEY parameters)
and measured tune shift values is shown in
Fig.~\ref{fig:2gev_tuneshift_sim} for the 2.1\unit{\gev} positron beam and in Fig.~\ref{fig:5gev_tuneshift_sim} for the 5.3\unit{\gev} beam.
Simulations based on the optimized SEY parameters agree at a
level better than 10\% with measurements
of tune shifts for all bunches in the train. Note that the tune shifts for different locations along the train
and for different beam energy and bunch current are in general dependent on distinct phenomena.
For example, the horizontal
tune shifts increase by about a factor of seven
when the bunch current is increased from 0.4 to 0.7~\unit{mA/bunch} at
2.1\unit{\gev}.
The model shows this dramatic effect to be dominated by cloud in the dipole sections
of the ring. On the other hand, the dipole and field-free regions contribute
comparably to the vertical tune shift at 0.4\unit{mA/bunch} and for the first ten bunches
of the train at 0.7\unit{mA/bunch}. While the vertical
tune shifts saturate at approximately 0.7\unit{kHz} for a bunch current of 0.4\unit{mA/bunch},
the dipole regions determine an approximately linear rise during the final
20 bunches at 0.7\unit{mA/bunch}, resulting in a final tune shift value about a factor of three higher
than at 0.4\unit{mA/bunch}.

Despite the 5.3\unit{\gev} bunch populations exceeding those in the 2.1\unit{\gev}
measurements by nearly a factor of ten,
the vertical tune shifts are less than a factor of two higher
than those at 2.1\unit{\gev}, a suppression which cannot be accounted for solely
by the beam stiffness. The dipole contributions show a threshold behavior
at 2\unit{mA/bunch} similar to that observed at 0.7\unit{mA/bunch} for the
2.1\unit{\gev} beam. The contribution of the field-free regions saturates
at a level of about 0.5--0.8\unit{kHz} at 5.3\unit{\gev}, roughly independently of bunch current
and similar to the level calculated by the model at 2.1\unit{\gev}
for a bunch current of 0.7\unit{mA/bunch}. At the higher bunch currents, the
vertical tune shifts begin to show some saturation, which is attributed to the
cloud behavior in the dipole regions of the ring. This saturation, or
reduction in tune shift increase, is particularly pronounced in the horizontal
tune shifts at 5.3\unit{\gev}, where again the dominant contributions are from
the dipole regions. In fact, the evolution in cloud shape along the train
results in increased suppression of the horizontal tune shifts for the higher
bunch currents, resulting in a decrease from a level of 3.5\unit{kHz},
similar to that measured in the vertical plane, to a value less than
0.5\unit{kHz} at 6\unit{mA/bunch}.

\begin{figure}[htbp]
\centering
\includegraphics[width=\columnwidth]{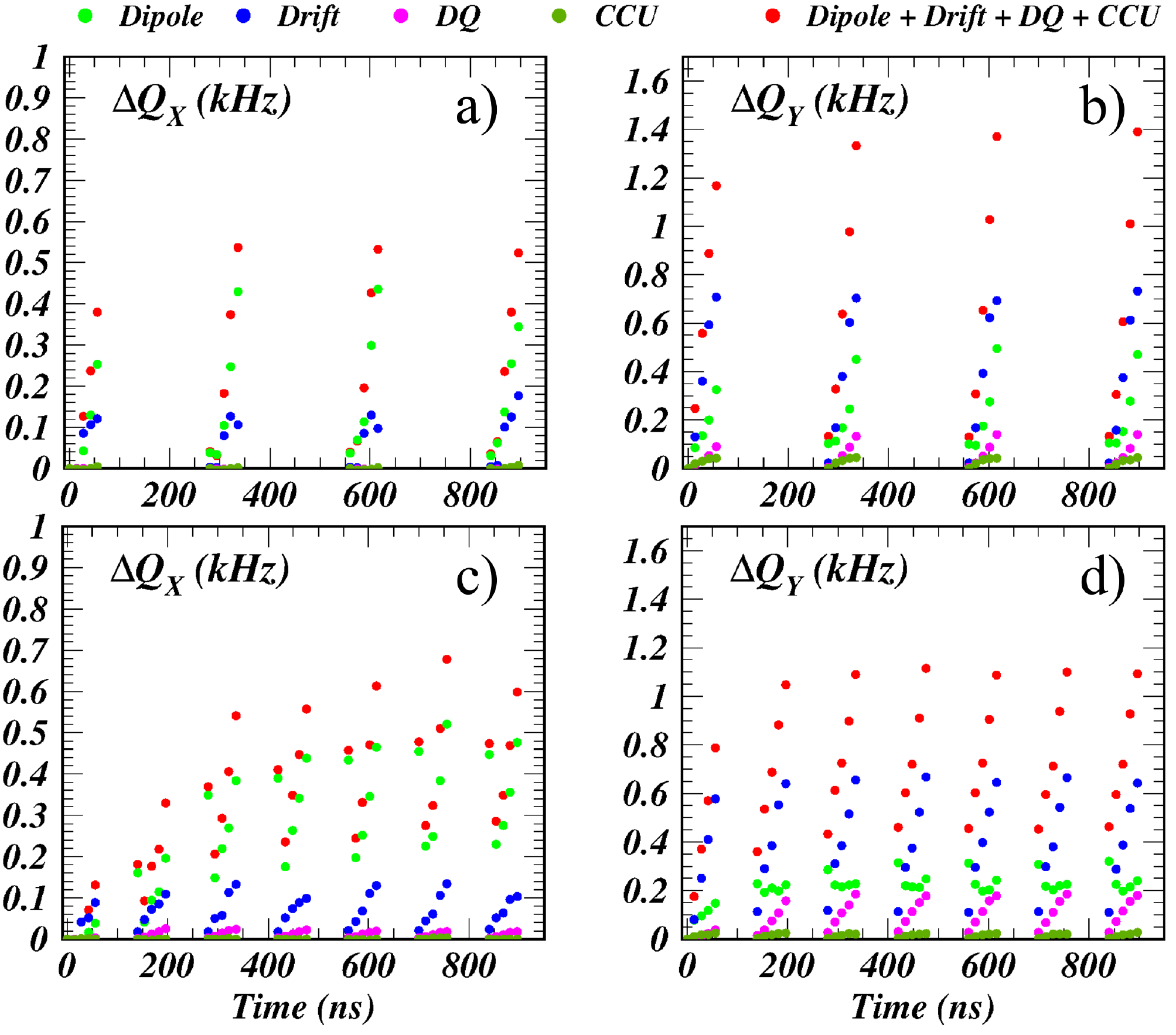}
   \caption{
     Modeled tune shifts for the 6.0\unit{\gev} CESR upgrade.
     The upper row shows the tune shifts in the a)~horizontal and b)~vertical planes  for
     the case of 9~trains of 5~bunches with bunch population $7.1\times10^{10}$.
     The lower row shows the tune shifts in the c)~horizontal and d)~vertical planes
     for the case of 18~trains of 5~bunches with bunch population $3.5\times10^{10}$.
     The bunch spacing within trains is 14\unit{ns} and the trains are equally spaced around the ring.
   }
   \label{fig:chessu}
\end{figure}

The validated model was employed to predict tune shifts for future
light source operation at 6\unit{\gev}
at the design beam current of 200\unit{mA}. 
The new
combined-function magnet regions and the compact permanent magnet undulator regions
were also included
in this study. While their ring occupancy fractions are low
(3.7\% and 2.9\%, respectively), the linear density per positron of
absorbed photons can be quite high, owing to the strong magnetic fields and the
locations of the magnets in the lattice. Nonetheless, the contributions
from these new magnets were found to be small, as shown in
Fig.~\ref{fig:chessu}.

We considered two configurations of 5-bunch trains equally spaced throughout the ring
giving the design value of 200\unit{mA} for the total beam current.
For 18 trains, or 9 trains with twice the bunch population,
the simulations show that the tune shifts
reach an equilibrium value
following the passage of just a few trains, which will be the level reached for the stored beam.
The maximum tune shift along the train is
found to be less than 2\unit{kHz}. While the dipole regions provide the
largest contribution to the horizontal tune shifts, the vertical tune shifts
show the field-free regions to dominate, owing to the short trains.
Since our measurements and modeling of tune shifts in the
pre-2019 high-current light source operation indicated that stable operation
was maintained with tune shifts of about 3\unit{kHz}, we conclude that the
tune shifts from electron cloud buildup will not prohibit reliable operation with positrons at the 
upgraded light source for the design beam current of 200\unit{mA}.

%

\section{Summary\label{sec:summary}}
We have obtained improved measurements of coherent betatron tune shifts 
along trains of positron bunches in the horizontal and vertical planes for 
bunch populations ranging from $0.64 \times 10^{10}$ to $9.6 \times 10^{10}$
at 2.1\unit{\gev} and 5.3\unit{\gev},
enabling advances in the predictive power of electron cloud buildup modeling.
Numerical simulation codes for photon tracking and photoelectron production
using a detailed model of the storage ring vacuum chamber
were employed to eliminate the ad hoc assumptions in electron production 
rates and kinematics endemic to prior buildup simulations.
A parametric model for secondary-yield processes was used in the electron cloud buildup
simulation to determine optimized parameters by fitting the modeled tune shift values
to the those measured.
Excellent agreement with the measurements was obtained for a wide variety of tune shift
patterns along the train,
allowing conclusions relating the tune shifts to various cloud buildup characteristics.
The model was then employed to predict the magnitude of tune shifts expected during future
operation of the Cornell Electron-positron Storage ring
as a high-brightness 6\unit{\gev} positron light source.
This study provides a high degree of confidence that stable
operation at 200\unit{mA} beam current can be achieved with either 9 or 18 trains of 5
positron bunches each. The generality and modularity of this modeling procedure addresses the
goal of the CESR Test Accelerator program to provide
design and diagnostic tools to other present and future accelerator facilities.

\section{Acknowledgments}
The authors wish to acknowledge important contributions from the technical 
staffs of the Wilson Laboratory. We thank Robert Meller for the design and implementation
of the tune tracker for betatron tune measurements.
This work is supported by the 
National Science Foundation under contracts no. PHY-0734867, no. PHY-1002467,
and by the U.S. Department of Energy under contracts
no. DE-FC02-08ER41538 and no. DE-SC0006505.



%

\end{document}